\documentclass[prd,aps,twocolumn,showpacs,floatfix,longbibliography,superscriptaddress]{revtex4-2}

\usepackage{mathrsfs}
\usepackage{rotating}
\usepackage{color}
\usepackage{graphicx}
\usepackage{array,booktabs}
\usepackage{amsmath,amssymb}
\usepackage{dcolumn}
\usepackage{bm}
\usepackage{hyperref}
\usepackage{braket}
\usepackage{pstricks}
\usepackage{version}
\usepackage{epstopdf}
\DeclareGraphicsExtensions{.eps,.ps}
\usepackage{ulem}
\usepackage{float}
\usepackage{subfigure}
\usepackage{setspace}
\usepackage[utf8]{inputenc}
\usepackage[T1]{fontenc}

\begin{document}

\title{Spin-Resolved Decay of Axion-Like Particles into Electron--Positron Pairs in Strong Electromagnetic Fields}

\author{Xiaodan Mao}
\affiliation{State Key Laboratory of Dark Matter Physics, Key Laboratory for Laser Plasmas (Ministry of Education) and School of Physics and Astronomy, Collaborative Innovation Center for IFSA (CICIFSA), Shanghai Jiao Tong University, Shanghai 200240, China}
\author{Yue-Yue Chen}
\email{yueyuechen@shnu.edu.cn}
\affiliation{Department of Physics, Shanghai Normal University, Shanghai 200234, China}
\author{Pei-Lun He}
\email{peilunhe@sjtu.edu.cn}
\affiliation{State Key Laboratory of Dark Matter Physics, Key Laboratory for Laser Plasmas (Ministry of Education) and School of Physics and Astronomy, Collaborative Innovation Center for IFSA (CICIFSA), Shanghai Jiao Tong University, Shanghai 200240, China}

\date{\today}
 
\begin{abstract}
We investigate spin-resolved decay of an axion-like particle (ALP) into an electron--positron pair in an intense laser field. Using the Baier–Katkov quasiclassical operator formalism in a constant crossed field, we derive a compact analytic rate retaining finite ALP-mass effects and the lepton spin degrees of freedom. In the massless limit, the spin-summed rate has the same weak- and strong-field asymptotic scalings as the corresponding
photon-induced pair-creation rate, while the pseudoscalar coupling induces distinct spin-resolved channels and spin correlations. A finite ALP mass reorganizes the spectrum across the vacuum threshold, producing purely field-induced pair creation below threshold and spin-dependent oscillatory modulations above threshold through the coherent interplay of vacuum and field-assisted contributions.
The entanglement of the produced pair reflects the dominant pair-creation mechanism. After the emission angles have been traced out, the reduced spin state near the vacuum threshold in weak fields is nearly maximally entangled and singlet-like. In the strong-field regime, it becomes triplet-like while retaining a concurrence approaching \(1/2\), whereas deep in the vacuum-dominated above-threshold regime it becomes triplet-like and separable. These results establish spin-resolved spectra and entanglement as characteristic signatures of finite-mass and threshold effects in strong-field ALP decay.
\end{abstract}

\maketitle


\section{Introduction}
\label{sec:intro}

Axion-like particles (ALPs) arise naturally in many extensions of the Standard Model, either as pseudo--Nambu--Goldstone bosons associated with spontaneously broken global symmetries~\cite{PhysRevD.16.1791, PhysRevLett.38.1440, PhysRevLett.43.103, SHIFMAN1980493, Zhitnitsky:1980tq, dine1981simple} or as generic light pseudoscalar states in ultraviolet completions such as string compactifications~\cite{witten1984some, arvanitaki2010string}. Their weak couplings to photons and fermions make them well-motivated dark matter candidates~\cite{dine1983not,abbott1983cosmological,preskill1983cosmology,marsh2016axion, IRASTORZA201889, di2020landscape, RevModPhys.93.015004, carenza2025axion} and promising targets for probing physics beyond the Standard Model in both astrophysical observations and laboratory experiments~\cite{pugnat2014search, ballou2015new, della2016pvlas, konaka1986search, davier1989unambiguous, bross1991search, abrahamyan2011search, battesti2018high}.
Complementary laboratory probes can also exploit virtual ALP effects. In particular, virtual ALP exchange can modify photon--photon interactions and photon propagation, giving rise to signatures such as ALP-mediated photon--photon scattering and vacuum birefringence~\cite{tommasini2009precision,dobrich2010axion,villalba2017polarization,evans2024searching}. 
Experimental searches for ALPs using stimulated resonant photon--photon collisions have also been reported~\cite{JHEP2025,Universe2026}.

Rapid advances in laser technology have enabled peak intensities approaching \(10^{22}\)--\(10^{23}\,\mathrm{W/cm^2}\)~\cite{Yoon:21}, opening new possibilities for probing ALPs in intense laser fields~\cite{gies2009strong}. In this regime, QED processes must be treated nonperturbatively with respect to the background field~\cite{di2012extremely,fedotov2023advances}. 
Prominent examples include nonlinear Compton scattering~\cite{nikishov1964quantum,ritus1970radiative,PhysRevA.83.022101,cole2018experimental,poder2018experimental} and nonlinear Breit--Wheeler pair production~\cite{reiss1962absorption,burke1997positron,dinu2016quantum,TangDillonKing2025,n1xm-66mq,seipt2020spin,tang2021pulse,tang2022fully,Ivanov2005,Jansen2016,Wistisen2020}, with recent studies of the latter addressing spin- and polarization-resolved rates, finite-pulse and pulse-envelope effects, and fully polarized processes in strong laser fields.
For ultrarelativistic particles in sufficiently strong laser fields, the characteristic formation scale of these processes is much smaller than the scale over which the background field varies. The local rate can then be approximated by the corresponding constant-crossed-field (CCF) rate~\cite{baier2005concept}, whose application along a slowly varying particle trajectory constitutes the locally constant field approximation (LCFA). The LCFA forms the basis of widely used semiclassical Monte Carlo simulations~\cite{elkina2011qed,ridgers2014modelling,green2015simla,gonoskov2015extended,zhuang2023laser} and also emerges as the leading term of a short-formation-length expansion in the Baier–Katkov quasiclassical operator method~\cite{baier1968quasiclassical,berestetskii2012quantum}.

Alongside extensive studies of the photon sector in strong-field QED, recent works have explored ALP production in intense laser backgrounds, in particular through nonlinear Compton-like emission from high-energy electrons~\cite{dillon2018alp,PhysRevD.111.055001,he2025semiclassical,chen2025spin}.
By crossing, the pseudoscalar ALP--electron interaction considered here also mediates laser-induced ALP decay into an electron--positron pair. This decay can occur even below the vacuum threshold, $m_a<2m_e$, where pair creation is enabled entirely by the background field. Compared with ALP emission, however, laser-induced ALP decay has received much less attention. Existing studies in oscillating plane waves~\cite{king2018electron} and CCFs~\cite{king2019axion} have focused mainly on spin-summed rates, leaving the spin correlations of the produced leptons unresolved.
The nonlinear Compton-like production and strong-field decay channels discussed above both depend directly on the ALP--electron coupling $g_{ae}$, which is subject to existing laboratory constraints on electrophilic pseudoscalars and ALPs~\cite{Eberhart2025,NA642021a,NA642021b}. If the ALPs are produced through nonlinear Compton-like emission mediated by $g_{ae}$, the production yield scales as $g_{ae}^2$ and the subsequent conversion probability likewise scales as $g_{ae}^2$ at leading order, so that the complete production--conversion event yield scales as $g_{ae}^4$. The decay rate derived below, however, is independent of how the incoming relativistic ALP is produced and can therefore be combined with a separately specified ALP source.

A useful comparison is provided by other pseudoscalar particles, most notably the neutral pion, whose properties and decay processes in strong electromagnetic backgrounds have also been studied~\cite{Ritus1969,ritus1985quantum}. At the level of a local pseudoscalar interaction, a particle described by the same effective vertex exhibits the same basic spin-dependent structure as the ALP considered here, with the mass and coupling replaced accordingly. The physical $\pi^0$, however, has $m_{\pi^0}\simeq135\,\mathrm{MeV}\simeq264\,m_e$, placing it far above the electron--positron threshold. By contrast, the present work focuses on ALP masses of order a few electron masses, including the below-threshold and near-threshold regions around $m_a=2m_e$. Thus, despite their common pseudoscalar structure, the physical $\pi^0$ and the ALPs considered here occupy very different mass and kinematic regimes.

A spin-resolved formulation is important both for identifying the characteristic spin-correlation structure of the pseudoscalar interaction and for practical implementation in strong-field simulations. In particular, the pseudoscalar coupling $\bar{\psi}\gamma^5\psi$ generates spin-correlation patterns distinct from those of the polarization-averaged photon-induced pair-creation channel, while semiclassical Monte Carlo simulations~\cite{chen2022electron,li2023strong,li2020polarized,dai2022photon,PhysRevD.110.012008,li2019ultrarelativistic,li2022helicity,chen2019polarized,li2020production,gong2021retrieving,PhysRevLett.129.035001,gong2023electron,PhysRevLett.131.225101,PhysRevLett.131.175101} require spin-resolved CCF rates as local kernels within the LCFA for ALP-induced pair production.

In this work, we study \(a\to e^+e^-\) in a CCF using the Baier--Katkov quasiclassical operator method and derive a compact analytic spin-resolved differential rate. The result retains finite ALP-mass effects and the spin degrees of freedom of the produced pair, giving a unified description across the vacuum threshold. For \(m_a<2m_e\), pair creation is purely field-induced, whereas above threshold the vacuum and field-assisted contributions combine coherently and generate spin-resolved oscillatory spectral modulations.
For a massless ALP, the spin-summed rate exhibits the same weak- and strong-field asymptotic scalings in the quantum nonlinearity parameter as the corresponding photon-induced pair-creation rate, up to an overall constant factor. The pseudoscalar coupling nevertheless produces qualitatively distinct spin-resolved channels and spin-correlation patterns. From the resulting rates, we construct the reduced two-particle spin density matrix and show that the spin correlations and entanglement are governed by the competition between vacuum and field-assisted decay contributions.

The remainder of this paper is organized as follows. In Sec.~\ref{sec:formalism} we introduce the interaction Lagrangian, establish the effective Frenet--Serret basis, and derive the general spin-resolved differential rate in a CCF. Section~\ref{sec:rate} presents the analytical evaluation of the decay rate, separately discussing the vacuum-forbidden and vacuum-allowed kinematic regions and deriving a unified Airy-function representation. In Sec.~\ref{sec:massless} we analyze in detail the massless limit, including the spin-resolved energy spectrum and the asymptotic weak- and strong-field behaviors. Section~\ref{sec:mass} is devoted to finite-mass effects, where we investigate the modification of the differential spectrum, the total decay rate, spin asymmetries across the \((\chi_a,\delta_m)\) parameter space, and the net positron polarization. In Sec.~\ref{sec:entanglement} we then extract the reduced two-qubit spin density matrix from the spin-resolved rate and study the quantum entanglement structure of the produced pair, with particular emphasis on the concurrence and the singlet-like versus triplet-like character of the spin correlations. Finally, Sec.~\ref{sec:conclusion} summarizes our findings and discusses possible applications in strong-field Monte Carlo simulations and future phenomenological studies.

\section{Formalism}
\label{sec:formalism}

\begin{figure}
\centering
\includegraphics[width=0.4\textwidth]{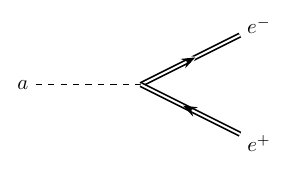}
\caption{
Feynman diagram for the decay of an ALP into an electron--positron pair in an intense laser field.
The dashed line represents the ALP, while the double lines denote the laser-dressed electron and positron propagating in the external field.
For \(m_a>2m_e\), the same final state is kinematically accessible already in vacuum; the decay rate then exhibits oscillatory structures associated with the vacuum--field crossover.
}
\label{FigFeyn}
\end{figure}

We follow the conventions of Ref.~\cite{he2025semiclassical} and summarize only the essential ingredients below. Throughout this work, we use natural units with \(\hbar=c=1\), unless stated otherwise.

\subsection{Interaction model}
\label{sec:lagrangian}

We describe the ALP--electron interaction by the pseudoscalar coupling
\begin{equation}
\mathcal{L}_I
=
- i\, g_{ae}\,\phi\,\bar{\psi}\gamma^5\psi\,,
\label{eq:Lagrangian}
\end{equation}
where \(\phi\) is the ALP field, $\psi$ the electron field, and \(g_{ae}\) the ALP--electron coupling constant.
Spacetime arguments are suppressed whenever no ambiguity arises.


We consider the decay of an ultrarelativistic ALP with energy \(\omega\), mass \(m_a\), and \(m_a^2/\omega^2\ll1\), in an intense laser background characterized by a large classical nonlinearity parameter \(a_0=g_eA_0/m_e\gg1\), where \(g_e\) denotes the QED coupling constant and \(A_0\) is the amplitude of the laser vector potential.
Excluding the extreme energy-sharing configurations in which one of the leptons carries almost all of the parent ALP energy, the produced electron and positron carry finite fractions of the parent energy and therefore satisfy $\varepsilon_\pm\gg m_e$. 
Such extreme configurations contribute negligibly to the total probability. The leptons' relativistic character is thus a physical feature of the high-energy regime considered here and simultaneously ensures the validity of the quasiclassical expansion.
In this regime, the formation length of strong-field pair production is much shorter than the characteristic scale over which the laser background varies~\cite{ritus1985quantum}, so that the local decay rate can be approximated by the corresponding CCF rate. Applying this local CCF rate along a slowly varying background constitutes the LCFA.
At fixed local field strength, the LCFA can be justified parametrically through the adiabatic limit \(\omega_L\to0\), for which \(a_0\to\infty\) while \(\chi_a\) remains fixed.
The remaining validity conditions are summarized in Ref.~\cite{he2025semiclassical}.

\subsection{Effective Frenet--Serret frame for the ALP}
\label{sec:FS}

To formulate the process in a quasiclassical form, we introduce an effective instantaneous Frenet--Serret-like orthonormal triad
\(\bigl(\mathbf{T},\,\mathbf{N},\,\mathbf{B}\bigr)\),
defined with respect to the ALP momentum and the local background field. The unit tangent vector is chosen along the ALP three-momentum, with
$\mathbf{T} = \frac{\mathbf{k}}{|\mathbf{k}|}\,$,
where \(\mathbf{k}\) denotes the corresponding momentum vector. We then define the component of the electric field parallel to the ALP direction and the effective transverse field as
$\boldsymbol{E}_{\mathrm{eff}}
=
\bigl(
\boldsymbol{E}_{\mathrm{phy}} \cdot \mathbf{T}
\bigr)\,\mathbf{T}\,$ and
$\boldsymbol{B}_{\mathrm{eff}}
=
\Bigl[
\Bigl(
\boldsymbol{E}_{\mathrm{phy}}
-\boldsymbol{E}_{\mathrm{eff}}
+\mathbf{T} \times \boldsymbol{B}_{\mathrm{phy}}
\Bigr)
\times \mathbf{T}
\Bigr]\,$,
where \(\boldsymbol{E}_{\mathrm{phy}}\) and \(\boldsymbol{B}_{\mathrm{phy}}\) are the physical electric and magnetic fields evaluated at the interaction point. The binormal direction is taken to be parallel to this effective transverse field, with
$\mathbf{B}
=
\frac{\boldsymbol{B}_{\mathrm{eff}}}
{|\boldsymbol{B}_{\mathrm{eff}}|}\,$,
and the normal vector is defined so as to complete the right-handed orthonormal triad,
$\mathbf{N}=\mathbf{B}\times\mathbf{T}\,$.

The strength of the background field relevant for the ALP decay is characterized by the quantum nonlinearity parameter
$\chi_a=
\frac{\omega\, g_e |\boldsymbol{B}_{\mathrm{eff}}|}{m_e^3}
\simeq
\frac{g_e}{m_e^3}
\sqrt{
\bigl(F_{\mu\nu} k^\nu\bigr)^2}$,
where \(F_{\mu\nu}\) is the background field-strength tensor, and we use the mostly-positive Minkowski metric.

\subsection{The spin-resolved decay rate in a CCF}
\label{sec:LCFA}
Although the ALP itself is electrically neutral, the short-formation-length condition underlying the LCFA is governed by the trajectories of the charged final-state particles. Applying the crossing relation for the spin labels [see Appendix~\ref{app:crossing_spin}] to the ALP-emission process derived in Ref.~\cite{he2025semiclassical}, we obtain the spin-resolved differential decay rate for $a\to e^+e^-$ in a CCF as
\begin{equation}
\frac{\mathrm{d}\Gamma}{\mathrm{d}\varepsilon_+}
= \frac{g_{ae}^2}{16\pi^3}\,
\frac{\varepsilon_{+}^2}{\omega}
\int \mathrm{d}\tau\,\mathrm{d}\alpha\,\mathrm{d}\beta\;
F(\tau,\alpha,\beta)\,
e^{-i(a\tau^3+b\tau)}\,,
\label{eq:DP}
\end{equation}
where the spin-dependent prefactor, evaluated up to \(O(m_e^2/\varepsilon_\pm^2)\), is given by
\begin{equation}
\begin{aligned}
F &\approx
\frac{1}{8}\,
\frac{\omega^2}{\varepsilon_-^{2}}
\Biggl\{
\left(
\frac{m_a^2}{\omega^2}\,
\frac{\varepsilon_-}{\varepsilon_+}
-\frac{1}{2}\,\omega_{\mathrm{eff}}^2\,\tau^2
\right)
\bigl(1-\boldsymbol{\zeta}_+\!\cdot\!\boldsymbol{\zeta}_-\bigr)
\\[4pt]
&\quad
+\,i\,\frac{m_e}{\varepsilon_+}\,
\omega_{\mathrm{eff}}\,\tau\;
\bigl(\boldsymbol{\zeta}_+-\boldsymbol{\zeta}_-\bigr)
\!\cdot\!\mathbf{B}
\\[4pt]
&\quad
+2\alpha^2\,
(\boldsymbol{\zeta}_+\!\cdot\!\mathbf{B})\,
(\boldsymbol{\zeta}_-\!\cdot\!\mathbf{B})
+2\beta^2\,
(\boldsymbol{\zeta}_+\!\cdot\!\mathbf{N})\,
(\boldsymbol{\zeta}_-\!\cdot\!\mathbf{N})
\\[4pt]
&\quad
-\frac{1}{2}\,\omega_{\mathrm{eff}}^2\,\tau^2\;
(\boldsymbol{\zeta}_+\!\cdot\!\mathbf{N})\,
(\boldsymbol{\zeta}_-\!\cdot\!\mathbf{N})
\\[4pt]
&\quad
+\left(
2\,\frac{m_a^2}{\omega^2}\,
\frac{\varepsilon_-}{\varepsilon_+}
-\omega_{\mathrm{eff}}^2\,\tau^2
-2\,\frac{m_e^2}{\varepsilon_+^{2}}
\right)
(\boldsymbol{\zeta}_+\!\cdot\!\mathbf{T})\,
(\boldsymbol{\zeta}_-\!\cdot\!\mathbf{T})
\Biggr\}.
\end{aligned}
\label{eq:Ftau}
\end{equation}
Here, \(\varepsilon_+\) and \(\varepsilon_-=\omega-\varepsilon_+\) are the positron and electron energies, respectively, and \(\boldsymbol{\zeta}_\pm\) denote the corresponding rest-frame spin vectors. The dimensionless variables \(\alpha\) and \(\beta\) parametrize the positron emission angle in the transverse \((\mathbf{N},\mathbf{B})\) plane, whereas \(\tau\) is the integration variable associated with the positron trajectory.

The phase in Eq.~\eqref{eq:DP} has the cubic structure characteristic of a CCF, with coefficients
\begin{equation}
\begin{aligned}
a &=
\frac{1}{24}
\frac{\varepsilon_+}{\varepsilon_-}
\omega \omega_{\mathrm{eff}}^{2},
\\[4pt]
b &=
\frac{1}{2}
\frac{\omega}{\varepsilon_-}
\frac{m_e^2}{\varepsilon_+}
\Delta
\left[
1+
\frac{\varepsilon_+^2}{m_e^2\Delta}
\bigl(\alpha^2+\beta^2\bigr)
\right].
\end{aligned}
\label{eq:ab}
\end{equation}
Here we have introduced the dimensionless mass ratio
\(\delta_m \equiv m_a/m_e\), the effective cyclotron frequency
\(\omega_{\mathrm{eff}}=g_e |\boldsymbol{B}_{\mathrm{eff}}|/\varepsilon_+\), which characterizes the local curvature of the charged-particle trajectory in the external field, and the threshold parameter
\begin{equation}
\Delta
=
1-
\frac{m_a^2}{m_e^2}
\frac{\varepsilon_-\varepsilon_+}{\omega^2}.
\label{eq:Delta}
\end{equation}
The parameter \(\Delta\) encodes the effect of the finite ALP mass and has a simple kinematic origin: for a fixed positron energy fraction \(\delta_+=\varepsilon_+/\omega\), vacuum decay into an electron--positron pair requires
\(\frac{m_a^2}{m_e^2}\delta_+(1-\delta_+) \geq 1\),
i.e., \(\Delta\leq 0\). 
This condition can never be met for \(m_a<2m_e\). For \(m_a>2m_e\), it holds only within the kinematic window \(\varepsilon_+\in\bigl(\varepsilon_+^{\min},\varepsilon_+^{\max}\bigr)\), where
\begin{equation}
\varepsilon_+^{\min,\max}=\frac{\omega}{2}\left(1\mp\sqrt{1-\frac{4m_e^2}{m_a^2}}\right).
\label{eq:window}
\end{equation}
Thus, at fixed energy partition, \(\Delta=0\) defines the boundary between the vacuum-forbidden and vacuum-allowed kinematic regions.

\begin{widetext}

\section{Analytical evaluation of the decay rate}
\label{sec:rate}

The integral in Eq.~\eqref{eq:DP} can be evaluated analytically, leading to qualitatively different forms in the two regimes identified below Eq.~\eqref{eq:Delta}. We first treat the vacuum-forbidden (\(\Delta>0\)) and vacuum-allowed (\(\Delta<0\)) kinematic regions in turn, then combine them into a unified Airy-function representation valid for arbitrary \(\Delta\), and finally verify that its zero-field limit reproduces the standard vacuum decay rate.

\subsection{\texorpdfstring{Vacuum-forbidden kinematic regime ($\Delta>0$)}{Vacuum-forbidden kinematic regime (Delta>0)}}
\label{sec:below}

For \(\Delta>0\), the coefficient \(b\) in Eq.~\eqref{eq:ab} is positive for all values of \(\rho=\sqrt{\alpha^2+\beta^2}\). Therefore, the \(\tau\) integral yields modified Bessel functions \(K_\nu\) via Eq.~\eqref{eq:intK}. The subsequent integration over the transverse variables \((\alpha,\beta)\) then proceeds as in Ref.~\cite{he2025semiclassical} for nonlinear Compton-like ALP emission. The resulting spin-resolved differential rate is
\begin{equation}
\begin{aligned}
\frac{\mathrm{d}\Gamma}{\mathrm{d}\varepsilon_+}
&= \frac{\sqrt{3}\,g_{ae}^2}{48\pi^2}\,
\frac{m_e^2}{\varepsilon_-\varepsilon_+}
\Biggl\{
\left(1-\boldsymbol{\zeta}_-\!\cdot\!\boldsymbol{\zeta}_+\right)
\left[
\Delta\,K_{\frac{2}{3}}(z_q)
+\frac{m_a^2}{2m_e^2}\,
\frac{\varepsilon_-\varepsilon_+}{\omega^2}
\int_{z_q}^{+\infty}\!\mathrm{d}t\;K_{\frac{1}{3}}(t)
\right]
\\[6pt]
&\quad
+
\bigl(\boldsymbol{\zeta}_+-\boldsymbol{\zeta}_-\bigr)
\!\cdot\!\mathbf{B}\,\sqrt{\Delta}\;K_{\frac{1}{3}}(z_q)
\\[6pt]
&\quad
+\frac{\Delta}{2}\,
(\boldsymbol{\zeta}_+\!\cdot\!\mathbf{N})\,
(\boldsymbol{\zeta}_-\!\cdot\!\mathbf{N})\,
\left[
3K_{\frac{2}{3}}(z_q)
-\int_{z_q}^{+\infty}\!\mathrm{d}t\;K_{\frac{1}{3}}(t)
\right]
\\[6pt]
&\quad
+\frac{\Delta}{2}\,
(\boldsymbol{\zeta}_+\!\cdot\!\mathbf{B})\,
(\boldsymbol{\zeta}_-\!\cdot\!\mathbf{B})\,
\left[
K_{\frac{2}{3}}(z_q)
-\int_{z_q}^{+\infty}\!\mathrm{d}t\;K_{\frac{1}{3}}(t)
\right]
\\[6pt]
&\quad
+\Delta\,
(\boldsymbol{\zeta}_+\!\cdot\!\mathbf{T})\,
(\boldsymbol{\zeta}_-\!\cdot\!\mathbf{T})\,
\left[
2K_{\frac{2}{3}}(z_q)
-\int_{z_q}^{+\infty}\!\mathrm{d}t\;K_{\frac{1}{3}}(t)
\right]
\Biggr\},
\end{aligned}
\label{eq:below}
\end{equation}
where
\begin{equation}
z_q=\frac{2}{3\chi_a}\,\frac{\omega^2}{\varepsilon_-\varepsilon_+}\,|\Delta|^{3/2},
\label{eq:zq}
\end{equation}
a definition that will also be used in the vacuum-allowed region.

\subsection{\texorpdfstring{Vacuum-allowed kinematic regime ($\Delta<0$)}{Vacuum-allowed kinematic regime (Delta < 0)}} \label{sec:above}

For $\Delta<0$, namely, for $\varepsilon_+$ within the kinematic window of Eq.~\eqref{eq:window}, the coefficient $b$ in Eq.~\eqref{eq:ab} changes sign at $\rho=\rho_0$, where $\rho_0=(m_e/\varepsilon_+)\sqrt{-\Delta}$. This divides the radial integral into an inner region, $\rho<\rho_0$, where $b<0$ and the $\tau$ integral yields ordinary Bessel functions $J_\nu$ through Eq.~\eqref{eq:intB}, and an outer region, $\rho>\rho_0$, where $b>0$ and modified Bessel functions $K_\nu$ appear as in the vacuum-forbidden regime. The full evaluation, detailed in Appendix~\ref{app:integral}, combines the contributions from both regions and yields the differential rate
\begin{equation}
\begin{aligned}
\frac{\mathrm{d}\Gamma}{\mathrm{d}\varepsilon_+}
&= \frac{g_{ae}^2}{48\pi}\frac{m_e^2}{\varepsilon_-\varepsilon_+}
\Biggl\{
\bigl(1-\boldsymbol{\zeta}_{-}\!\cdot\!\boldsymbol{\zeta}_{+}\bigr)
\left[
\Delta\,\mathcal{J}_{-}(z_q)
+\frac{m_a^2}{2m_e^2}\frac{\varepsilon_-\varepsilon_+}{\omega^2}
\left(
3-\int_{z_q}^{+\infty}\!\mathrm{d}t\,\mathcal{J}_{+}(t)
\right)
\right]
\\[6pt]
&\quad
+\bigl(\boldsymbol{\zeta}_{+}-\boldsymbol{\zeta}_{-}\bigr)\!\cdot\!\mathbf{B}\sqrt{-\Delta}\,\mathcal{J}_{+}(z_q)
\\[6pt]
&\quad
-\frac{\Delta}{2}
(\boldsymbol{\zeta}_{+}\!\cdot\!\mathbf{N})
(\boldsymbol{\zeta}_{-}\!\cdot\!\mathbf{N})
\left[
3-\int_{z_q}^{+\infty}\!\mathrm{d}t\,\mathcal{J}_{+}(t)
-3\,\mathcal{J}_{-}(z_q)
\right]
\\[6pt]
&\quad
-\frac{\Delta}{2}
(\boldsymbol{\zeta}_{+}\!\cdot\!\mathbf{B})
(\boldsymbol{\zeta}_{-}\!\cdot\!\mathbf{B})
\left[
3-\int_{z_q}^{+\infty}\!\mathrm{d}t\,\mathcal{J}_{+}(t)
-\mathcal{J}_{-}(z_q)
\right]
\\[6pt]
&\quad
+\Delta
(\boldsymbol{\zeta}_{+}\!\cdot\!\mathbf{T})
(\boldsymbol{\zeta}_{-}\!\cdot\!\mathbf{T})
\left[
2\,\mathcal{J}_{-}(z_q)
+\int_{z_q}^{+\infty}\!\mathrm{d}t\,\mathcal{J}_{+}(t)
-3
\right]
\Biggr\},
\end{aligned}
\label{eq:above}
\end{equation}
where $z_q$ is given by Eq.~\eqref{eq:zq}, and the combinations of Bessel functions $\mathcal{J}_\pm$ are defined in Appendix~\ref{app:Jpm}.

Equation~\eqref{eq:above} has the same spin-dependent tensor structure as the vacuum-forbidden result in Eq.~\eqref{eq:below}, but the exponentially decaying modified Bessel functions $K_\nu(z)$ are replaced by the oscillatory combinations $\mathcal{J}_\pm(z)$. This change in analytic structure reflects the opening of the vacuum decay channel at $\Delta=0$ and admits a simple saddle-point interpretation, reminiscent of St\"uckelberg interference in pair-production processes~\cite{dumlu2010stokes,dumlu2011interference,akkermans2012ramsey}. For fixed transverse variables, the formation-time phase in Eq.~\eqref{eq:DP} has the form $\Phi(\tau)=a\tau^3+b\tau$, with $a>0$, and its stationary points satisfy $3a\tau_s^2+b=0$.

For $b>0$, the two saddle points lie on the imaginary axis at $\tau_s=\pm i\sqrt{b/(3a)}$. Under the appropriate deformation of the integration contour, only the saddle at $\tau_s=-i\sqrt{b/(3a)}$ contributes, yielding an exponentially suppressed contribution described by modified Bessel functions. This situation applies throughout the transverse phase space when $\Delta>0$.

When $\Delta<0$, a finite region of transverse phase space opens in which $b<0$. In this region, the two saddle points move onto the real axis, $\tau_s=\pm\sqrt{-b/(3a)}$, and both contribute coherently to the decay probability. Their phase difference is
\(
\Delta\Phi=4a\left(\frac{-b}{3a}\right)^{3/2}.
\)
After integration over the transverse variables, this real-saddle interference is encoded in the variable $z_q$ and gives rise to the oscillatory combinations $\mathcal{J}_\pm$ in Eq.~\eqref{eq:above}.

The physical origin of these oscillations can be made more transparent in the equivalent Volkov-state description, in which the interaction with the background is resummed to all orders in $a_0$~\cite{ritus1985quantum,di2012extremely}. In a formal expansion of the dressed amplitude in powers of the external field, the zeroth-order term reproduces the vacuum amplitude for $a\to e^+e^-$, which becomes kinematically allowed above threshold, while higher-order terms describe background-assisted contributions. Their coherent superposition generates interference terms whenever both contributions are present. A related amplitude-level organization has been employed in the study of muon decay in an external field~\cite{823w-2g4b}. In the present Baier--Katkov formulation, the vacuum and background-assisted contributions are not separated explicitly at the amplitude level. Instead, the same interference is encoded in the phase difference between the two real formation-time saddles, producing the characteristic $z_q$-dependent oscillatory pattern.

\subsection{Unified Airy-function representation} \label{sec:Airy}

The two kinematic regimes derived above can be combined into a single compact expression using the standard relations between Airy functions and Bessel functions of orders \(1/3\) and \(2/3\) (see Appendix~\ref{app:Airy}). Rewriting Eqs.~\eqref{eq:below} and \eqref{eq:above} in this form yields the differential rate valid for arbitrary \(\Delta\):
\begin{equation}
\begin{aligned}
\frac{\mathrm{d}\Gamma}{\mathrm{d}\varepsilon_{+}}
&= \frac{g_{ae}^2}{16\pi}\frac{m_e^2}{\varepsilon_{+}\varepsilon_{-}}
\Biggl\{
\left(
-\Delta\,\frac{\mathrm{Ai}'(u)}{u}
+\frac{1}{2}\frac{m_a^2}{m_e^2}\frac{\varepsilon_{-}\varepsilon_{+}}{\omega^2}
\int_{u}^{+\infty}\!\mathrm{d}t\,\mathrm{Ai}(t)
\right)
\bigl(1-\boldsymbol{\zeta}_{-}\!\cdot\!\boldsymbol{\zeta}_{+}\bigr)
\\[6pt]
&\quad
+\sqrt{\frac{\Delta}{u}}\,\mathrm{Ai}(u)\,
\bigl(\boldsymbol{\zeta}_{+}-\boldsymbol{\zeta}_{-}\bigr)\!\cdot\!\mathbf{B}
\\[6pt]
&\quad
-\frac{\Delta}{2}
\left(
\frac{3\mathrm{Ai}'(u)}{u}
+\int_{u}^{+\infty}\!\mathrm{d}t\,\mathrm{Ai}(t)
\right)
(\boldsymbol{\zeta}_{-}\!\cdot\!\mathbf{N})
(\boldsymbol{\zeta}_{+}\!\cdot\!\mathbf{N})
\\[6pt]
&\quad
-\frac{\Delta}{2}
\left(
\frac{\mathrm{Ai}'(u)}{u}
+\int_{u}^{+\infty}\!\mathrm{d}t\,\mathrm{Ai}(t)
\right)
(\boldsymbol{\zeta}_{-}\!\cdot\!\mathbf{B})
(\boldsymbol{\zeta}_{+}\!\cdot\!\mathbf{B})
\\[6pt]
&\quad
-\Delta
\left(
2\,\frac{\mathrm{Ai}'(u)}{u}
+\int_{u}^{+\infty}\!\mathrm{d}t\,\mathrm{Ai}(t)
\right)
(\boldsymbol{\zeta}_{-}\!\cdot\!\mathbf{T})
(\boldsymbol{\zeta}_{+}\!\cdot\!\mathbf{T})
\Biggr\},
\end{aligned}
\label{eq:Airy}
\end{equation}
where
\(
u=\Delta\left(\frac{\omega^2}{\chi_a\varepsilon_-\varepsilon_+}\right)^{2/3}.
\)
Since
\(
\frac{\Delta}{u}=\left(\frac{\chi_a\varepsilon_-\varepsilon_+}{\omega^2}\right)^{2/3},
\)
the ratio \(\Delta/u\) is real and positive on both sides of the threshold. At \(\Delta=0\), the corresponding factor in Eq.~\eqref{eq:Airy} is understood by continuity. Equation~\eqref{eq:Airy} thus provides a single uniform representation across the threshold, reducing to Eq.~\eqref{eq:below} for \(\Delta>0\) and to Eq.~\eqref{eq:above} for \(\Delta<0\).

Equation~\eqref{eq:Airy} is one of the main results of this work. Its compact form is particularly convenient for numerical and Monte Carlo implementations, since the \(\Delta>0\) and \(\Delta<0\) regions need not be treated separately. As a consistency check, our result, summed over the spins of the produced electron and positron, agrees with the spin-summed CCF expression obtained from Volkov wave functions in Ref.~\cite{king2019axion}.

In addition, Eq.~\eqref{eq:Airy} exhibits a discrete symmetry under the combined exchange and sign reversal of the electron and positron spin vectors,
\begin{equation}
\boldsymbol{\zeta}_{+}\rightarrow-\boldsymbol{\zeta}_{-},
\qquad
\boldsymbol{\zeta}_{-}\rightarrow-\boldsymbol{\zeta}_{+}.
\label{eq:SpinSym}
\end{equation}
Under this transformation, the differential rate remains invariant.

\subsection{Vacuum decay limit and weak-field interpretation} \label{sec:vacuum}

We now consider the formal zero-field limit $\chi_a\to0$ of the unified Airy expression in Eq.~\eqref{eq:Airy}. As discussed in Sec.~\ref{sec:lagrangian}, this limit remains compatible with the LCFA by taking $\omega_L\to0$ sufficiently rapidly to maintain $a_0\gg1$. In this limit, Eq.~\eqref{eq:Airy} reduces to the spin-resolved vacuum decay spectrum
\begin{equation}
\begin{aligned}
\frac{\mathrm{d}\Gamma}{\mathrm{d}\varepsilon_{+}}
&= \frac{g_{ae}^2}{16\pi}\frac{m_e^2}{\varepsilon_{+}\varepsilon_{-}}
\Biggl\{
\left(
\frac{1}{2}\frac{m_a^2}{m_e^2}\frac{\varepsilon_{-}\varepsilon_{+}}{\omega^2}
\right)
\bigl(1-\boldsymbol{\zeta}_{-}\!\cdot\!\boldsymbol{\zeta}_{+}\bigr)
\\[6pt]
&\quad
-\frac{\Delta}{2}
(\boldsymbol{\zeta}_{-}\!\cdot\!\mathbf{N})
(\boldsymbol{\zeta}_{+}\!\cdot\!\mathbf{N})
-\frac{\Delta}{2}
(\boldsymbol{\zeta}_{-}\!\cdot\!\mathbf{B})
(\boldsymbol{\zeta}_{+}\!\cdot\!\mathbf{B})
-\Delta
(\boldsymbol{\zeta}_{-}\!\cdot\!\mathbf{T})
(\boldsymbol{\zeta}_{+}\!\cdot\!\mathbf{T})
\Biggr\},
\end{aligned}
\label{eq:vac_differential}
\end{equation}
where the positron energy is restricted to the kinematic window $\varepsilon_+\in\bigl(\varepsilon_+^{\min},\varepsilon_+^{\max}\bigr)$ given in Eq.~\eqref{eq:window}. After summing over the final-state spins and integrating over this kinematic window, one obtains
\begin{equation}
\Gamma_{\mathrm{vac}}
= \frac{g_{ae}^2}{8\pi}\frac{m_a^2}{\omega}\sqrt{1-\frac{4m_e^2}{m_a^2}},
\label{eq:vac_total}
\end{equation}
which is the standard vacuum decay rate in the laboratory frame. The factor $m_a/\omega$ is the inverse Lorentz factor and accounts for time dilation between the ALP rest frame and the laboratory frame. The resulting expression therefore reproduces the correct zero-field limit in the above-threshold regime, expressed in the same ultrarelativistic spin basis.
\end{widetext}


\section{Field-induced decay of massless ALPs}
\label{sec:massless}

In this section, we examine ALP-induced pair production in the massless limit, \(m_a=0\), corresponding to \(\Delta=1\). This limit is phenomenologically relevant for light ALPs with masses far below the pair-production threshold: the vacuum decay channel is kinematically forbidden, and pair creation is entirely induced by the external field. The rate is then given by the below-threshold result, Eq.~\eqref{eq:below}, with the argument of the modified Bessel functions reducing to \(z_q=\frac{2}{3\chi_a\,\delta_+(1-\delta_+)}\). In the numerical results presented below, we take $\omega=1~\mathrm{GeV}$ as a representative ultrarelativistic benchmark, consistent with typical GeV-scale ALP energies accessible through nonlinear Compton-like production mechanisms in intense laser fields~\cite{he2025semiclassical,chen2025spin}. At fixed $\chi_a$, $\delta_m$, and energy fraction $\delta_+$, the dimensionless spectral and spin structures are independent of $\omega$, while the dimensional rates scale as $d\Gamma/d\varepsilon_+\propto\omega^{-2}$ and $\Gamma\propto\omega^{-1}$. For a fixed physical field, however, $\chi_a\propto\omega$, so changing the ALP energy also changes the strong-field regime.

\subsection{Spin-resolved energy spectrum}
\label{sec:spectrum}

\begin{figure*}
\centering
\includegraphics[width=1\textwidth]{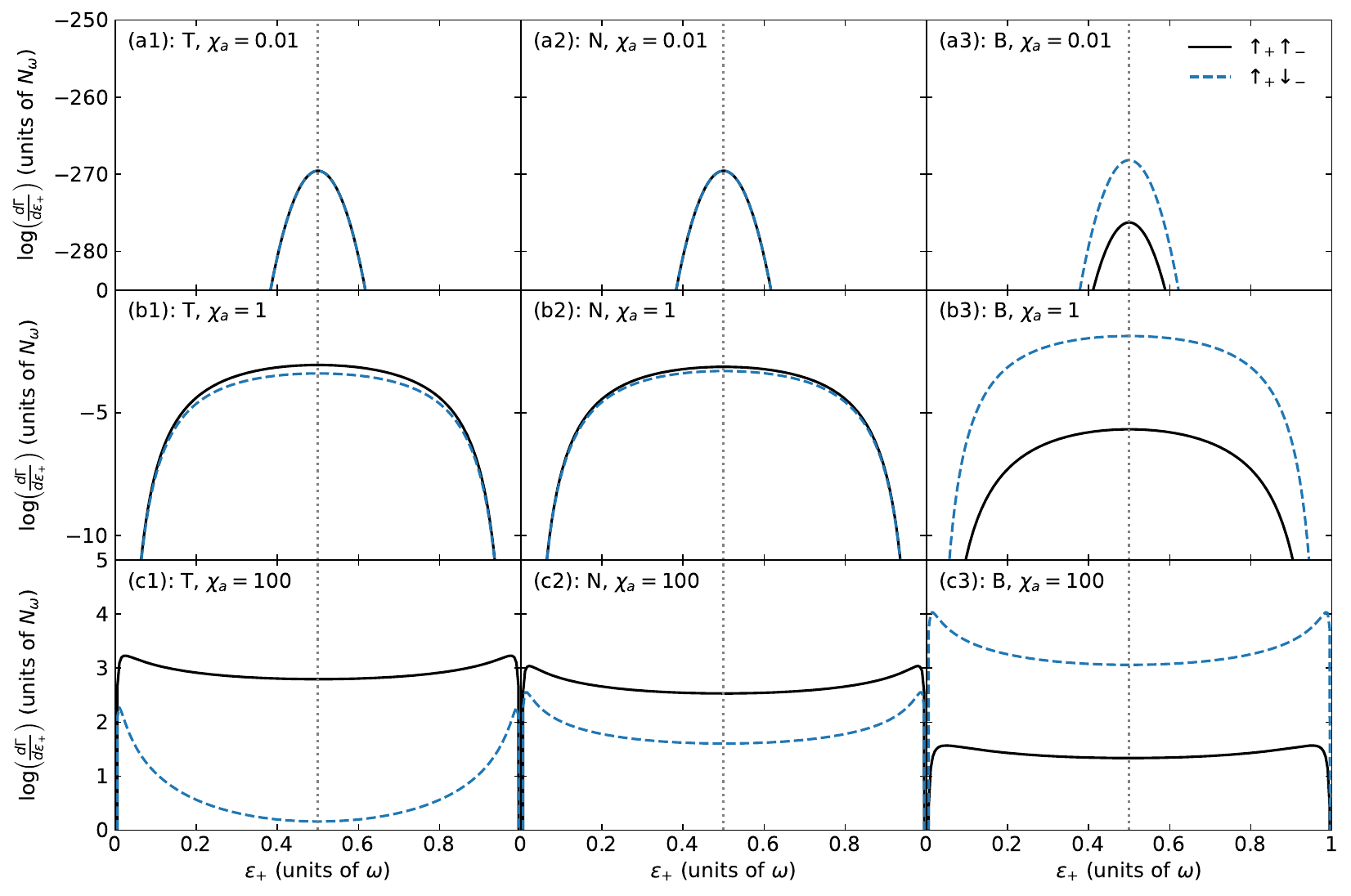}
\caption{Logarithm of the spin-resolved differential decay rate \(\mathrm{d}\Gamma/\mathrm{d}\varepsilon_+\), in units of \(N_\omega=\frac{g_{ae}^2}{16\pi}\frac{m_e^2}{\omega^2}\), as a function of the positron energy fraction \(\delta_+=\varepsilon_+/\omega\) for different spin configurations. The top, middle, and bottom rows correspond to \(\chi_a=0.01\), \(1\), and \(100\), representing the weak-field, intermediate-field, and asymptotic strong-field regimes, respectively. In the left, middle, and right columns, the positron spin is fixed along the positive \(\mathbf{T}\), \(\mathbf{N}\), and \(\mathbf{B}\) directions, respectively, while the electron spin is chosen to be either parallel (solid black curves) or antiparallel (dashed blue curves) to the positron spin. The ALP energy is \(\omega=1~\mathrm{GeV}\), and its normalized mass is \(\delta_m=m_a/m_e=0\).}
\label{FigdGamma}
\end{figure*}

Figure~\ref{FigdGamma} displays the spin-resolved differential rate \(\mathrm{d}\Gamma/\mathrm{d}\varepsilon_+\) as a function of \(\delta_+=\varepsilon_+/\omega\) for \(\chi_a=0.01\), \(1\), and \(100\), with the positron spin quantized along the \(\mathbf{T}\), \(\mathbf{N}\), and \(\mathbf{B}\) directions from left to right. The shape of the spin-resolved energy distribution is controlled by \(\chi_a\) through the Bessel-function argument \(z_q\).

In the weak-field regime, \(\chi_a\ll1\), one has \(z_q\gg1\) throughout the entire phase space. The Bessel functions therefore reduce to their large-argument asymptotic form, \(K_\nu(z_q)\sim e^{-z_q}\), and the rate is dominated by the exponential factor. Since \(z_q\) is minimized at \(\delta_+=1/2\), the spectrum is sharply peaked at equal energy sharing between the electron and positron.
This is the characteristic signature of the tunneling regime, in which pair creation proceeds nonperturbatively through the external field and only the most favorable kinematic configuration survives the exponential suppression. Expanding around the midpoint,
\begin{equation}
\delta_+ = \frac12 + \epsilon, \qquad |\epsilon| \ll 1,
\end{equation}
one finds
\[
z_q \approx \frac{8}{3 \chi_a} \left( 1 + 4 \epsilon^2 \right),
\]
and hence
\begin{equation}
e^{-z_q} \sim
\exp\!\left(-\frac{8}{3\chi_a}\right)
\exp\!\left(-\frac{32 \epsilon^2}{3\chi_a}\right),
\end{equation}
corresponding to an approximately Gaussian dependence with a width of order \(\sqrt{\chi_a}\). The spectrum therefore narrows as \(\chi_a\) decreases [Figs.~\ref{FigdGamma}(a1--a3)] and broadens as \(\chi_a\) approaches unity and the exponential suppression weakens [Figs.~\ref{FigdGamma}(b1--b3)].

In the strong-field regime, \(\chi_a\gg1\), the condition \(z_q\leq1\) is satisfied whenever \(\delta_+(1-\delta_+)\geq\frac{2}{3\chi_a}\), i.e., over most of the phase space. In this region, the Bessel functions take their small-argument power-law form and the tunneling suppression disappears. Only near the endpoints, \(\delta_+\to0\) and \(\delta_+\to1\), does \(z_q\) grow again, cutting off the endpoint enhancement within regions of width \(\sim1/\chi_a\). The spectrum therefore develops power-law peaks near both endpoints, reflecting a pronounced preference for asymmetric energy sharing between the electron and positron [Figs.~\ref{FigdGamma}(c1--c3)].

Regarding the spin correlations, the \(\mathbf{B}\) direction exhibits the most robust spin preference, with the preferred spin orientations remaining unchanged over the full range of \(\chi_a\). The electron preferentially emerges with spin antiparallel to \(\mathbf{B}\), whereas the positron tends to be polarized parallel to \(\mathbf{B}\). This persistent preference follows directly from the single-spin term proportional to $(\boldsymbol{\zeta}_{+}-\boldsymbol{\zeta}_{-})\cdot\mathbf{B}$ in the spin-resolved rate and reflects the pseudoscalar structure of the interaction.

The correlations along \(\mathbf{T}\) and \(\mathbf{N}\), by contrast, depend strongly on \(\chi_a\). In the tunneling regime, \(\chi_a\ll1\), the spectra for parallel and antiparallel spin projections are nearly identical along both axes [Figs.~\ref{FigdGamma}(a1) and (a2)], indicating that the spin correlations along \(\mathbf{T}\) and \(\mathbf{N}\) are strongly suppressed; see also Eq.~\eqref{eq:weak_ALP}. As \(\chi_a\) increases, the interaction with the background field generates visible correlations along \(\mathbf T\) and \(\mathbf N\). For \(\chi_a\gg1\) [Figs.~\ref{FigdGamma}(c1) and (c2)], the electron tends to align its spin with that of the positron along either the \(\mathbf T\) or \(\mathbf N\) axis. This preference for parallel configurations is stronger along \(\mathbf{T}\) than along \(\mathbf{N}\), as seen from the larger splitting between \(\mathrm{d}\Gamma_{\uparrow_+\uparrow_-}\) and \(\mathrm{d}\Gamma_{\uparrow_+\downarrow_-}\) for quantization along \(\mathbf{T}\); see also Eq.~\eqref{eq:strong_ALP}. 
The background field therefore both induces pair creation and imprints a characteristic pattern of spin correlations on the produced pair, an effect essentially absent in the tunneling regime.

\subsection{Asymptotic limits}
\label{sec:asymptotic}

Integrating the differential rate in Eq.~\eqref{eq:below} over the positron energy yields simple closed-form expressions in the weak- and strong-field limits. In both regimes, the spin-summed ALP-induced and photon-induced rates share the same parametric dependence on the field strength, with exponential tunneling suppression for \(\chi\ll1\) and a \(\chi^{2/3}\) power law for \(\chi\gg1\), reflecting the universal structure of strong-field pair creation in external fields.

\subsubsection*{\texorpdfstring{Weak-field limit \(\chi_a\ll1\)}{Weak-field limit}}

In the tunneling regime, the dominant contribution comes from the symmetric energy configuration, \(\delta_+=1/2\), and the ALP decay rate takes the form
\begin{equation}
\begin{aligned}
\Gamma \approx
\frac{\sqrt{6}\,g_{ae}^2}{128\pi}\,
\frac{m_e^2}{\omega}\,
\chi_a\,
\exp\!\left(-\frac{8}{3\chi_a}\right)
\Biggl[
1 &- (\boldsymbol{\zeta}_+\!\cdot\!\mathbf{B})\,
      (\boldsymbol{\zeta}_-\!\cdot\!\mathbf{B})
\\
&+ (\boldsymbol{\zeta}_+-\boldsymbol{\zeta}_-)
   \!\cdot\!\mathbf{B}
\Biggr].
\end{aligned}
\label{eq:weak_ALP}
\end{equation}
The spin dependence in Eq.~\eqref{eq:weak_ALP} is remarkably simple: only the projections of the electron and positron polarizations onto the local effective magnetic-field direction \(\mathbf{B}\) enter, while the spin-correlation components along \(\mathbf T\) and \(\mathbf N\) drop out entirely. This is consistent with the spin-resolved spectra in Figs.~\ref{FigdGamma}(a1--a3).

For comparison, the corresponding polarization-averaged nonlinear Breit--Wheeler pair-production rate~\cite{berestetskii2012quantum,chen2022electron} in the same limit reads
\begin{equation}
\begin{aligned}
\Gamma_\gamma
&\approx
\sqrt{\frac{3}{2}}\;\frac{g_e^2}{128\pi}\,
\frac{m_e^2}{\omega}\,
\chi_\gamma\,
\exp\!\left(-\frac{8}{3\chi_\gamma}\right)
\\
&\quad\times
\Biggl[
\frac{3}{2}
-\frac{1}{2}\,\boldsymbol{\zeta}_-\!\cdot\!\boldsymbol{\zeta}_+
+\bigl(\boldsymbol{\zeta}_+-\boldsymbol{\zeta}_-\bigr)
\!\cdot\!\mathbf{B}
\\
&\qquad
+(\boldsymbol{\zeta}_-\!\cdot\!\mathbf{T})\,
 (\boldsymbol{\zeta}_+\!\cdot\!\mathbf{T})
\Biggr],
\end{aligned}
\label{eq:weak_photon}
\end{equation}
where \(\chi_\gamma=g_e\sqrt{(F_{\mu\nu}k^\nu)^2}/m_e^3\) denotes the photon quantum nonlinearity parameter. 
The comparison throughout this section refers to the polarization-averaged nonlinear Breit--Wheeler channel. For a polarized incident photon, the pair-spin correlations depend additionally on the photon Stokes parameters, and the corresponding polarization-resolved rate should be used for comparison~\cite{seipt2020spin,TangDillonKing2025}.
The two rates contain the same nonperturbative factor, \(\exp[-8/(3\chi)]\), and the same single-spin asymmetries along \(\mathbf{B}\). Thus, both processes favor positrons polarized parallel to \(\mathbf{B}\) and electrons polarized antiparallel to \(\mathbf{B}\). However, the spin--spin correlation structures of the two channels differ.

\subsubsection*{\texorpdfstring{Strong-field limit \(\chi_a\gg1\)}{Strong-field limit}}

In the opposite regime, the ALP-induced rate exhibits power-law scaling:
\begin{equation}
\begin{aligned}
\Gamma &\approx 0.133\;\frac{g_{ae}^2}{4\pi}\,
\frac{m_e^2}{\omega}\,\chi_a^{2/3}
\Biggl[
1 + (\boldsymbol{\zeta}_+\!\cdot\!\mathbf{T})\,
     (\boldsymbol{\zeta}_-\!\cdot\!\mathbf{T})
\\
&\qquad
+\frac{1}{2}\,
(\boldsymbol{\zeta}_+\!\cdot\!\mathbf{N})\,
(\boldsymbol{\zeta}_-\!\cdot\!\mathbf{N})
-\frac{1}{2}\,
(\boldsymbol{\zeta}_+\!\cdot\!\mathbf{B})\,
(\boldsymbol{\zeta}_-\!\cdot\!\mathbf{B})
\Biggr],
\end{aligned}
\label{eq:strong_ALP}
\end{equation}
while the corresponding polarization-averaged nonlinear Breit--Wheeler pair-production rate~\cite{berestetskii2012quantum,chen2022electron} is
\begin{equation}
\begin{aligned}
\Gamma_\gamma
&\approx 0.133\;\frac{g_e^2}{4\pi}\,
\frac{m_e^2}{\omega}\,\chi_\gamma^{2/3}
\biggl[
\frac{5}{7}
-\frac{3}{7}\,
(\boldsymbol{\zeta}_+\!\cdot\!\mathbf{T})\,
(\boldsymbol{\zeta}_-\!\cdot\!\mathbf{T})
\\
&\qquad
-\frac{2}{7}\,
\boldsymbol{\zeta}_+\!\cdot\!\boldsymbol{\zeta}_-
\biggr].
\end{aligned}
\label{eq:strong_photon}
\end{equation}
Both channels display the universal $\chi^{2/3}$ scaling, but their spin-correlation patterns are qualitatively distinct: the ALP rate favors parallel electron--positron spin configurations when the quantization axis is chosen along $\mathbf{T}$ or $\mathbf{N}$, and antiparallel configurations when it is chosen along $\mathbf{B}$. By contrast, the polarization-averaged photon rate favors antiparallel configurations for all quantization axes.


\section{Finite-mass effects}
\label{sec:mass}

\begin{figure*}
\centering
\includegraphics[width=1\textwidth]{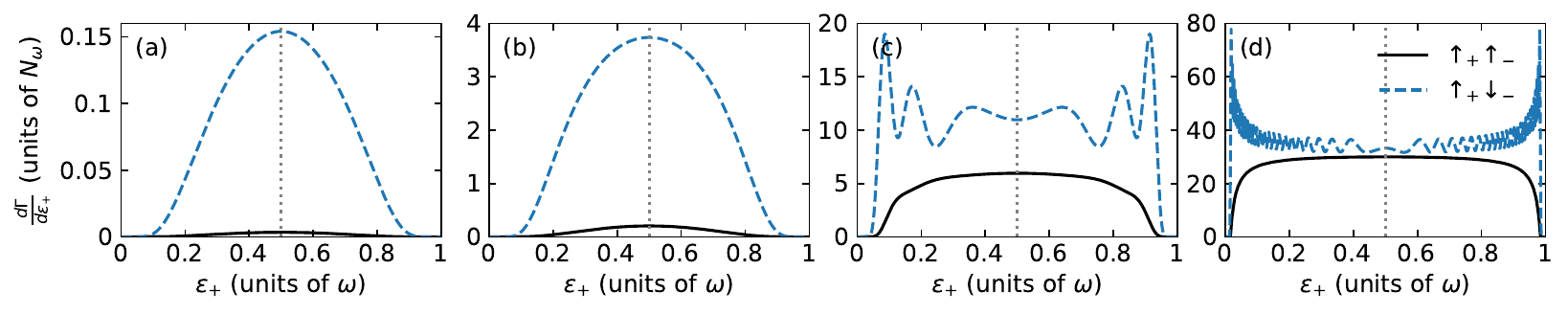}
\caption{%
Spin-resolved differential decay rate \(\mathrm{d}\Gamma/\mathrm{d}\varepsilon_+\) in units of \(N_\omega\) as a function of the positron energy fraction \(\delta_+=\varepsilon_+/\omega\), at fixed \(\chi_a=1\), with the positron spin fixed along the positive \(\mathbf B\) direction. Solid black curves denote the configuration in which the electron spin is parallel to the positron spin, while dashed blue curves denote the antiparallel configuration. Panels~(a)--(d) correspond to the normalized ALP masses \(\delta_m=m_a/m_e=0,\,2,\,4,\) and \(8\), respectively.}
\label{FigMass}
\end{figure*}

Having investigated the properties of the decay rate in the massless limit, we now turn to the role of a finite ALP mass. Varying the mass across the pair-production threshold, \(\delta_m=2\), qualitatively changes the spin-resolved spectra.

\subsection{Spin-resolved spectrum, total rate, and spin asymmetry}
\label{sec:massspectrum}

\begin{figure*}
\centering
\includegraphics[width=\textwidth]{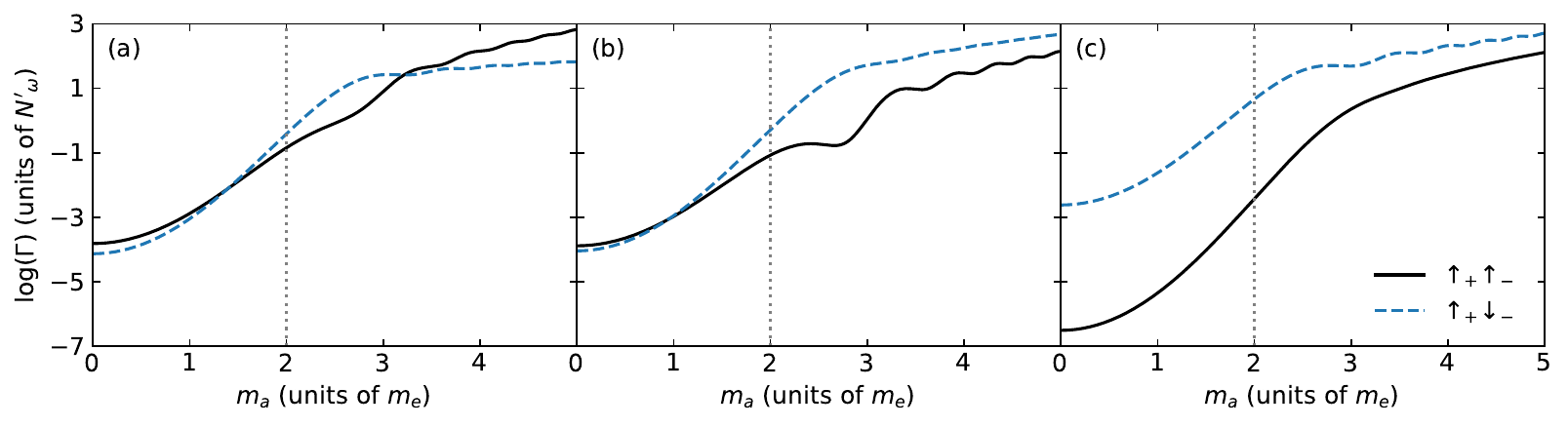}
\caption{
Logarithm of the spin-resolved total pair-production rate~\(\Gamma\) in units of \(N'_\omega=\frac{g_{ae}^2}{16\pi}\,\frac{m_e^2}{\omega}\) as a function of the ALP mass, at fixed \(\chi_a=1\).
Panels~(a)--(c) correspond to the positron spin fixed along the positive \(\mathbf{T}\), \(\mathbf{N}\), and \(\mathbf{B}\) directions, respectively, while the electron spin is chosen either parallel (solid black curves) or antiparallel (dashed blue curves) to the positron spin.
}
\label{FigMassTotal}
\end{figure*}

Figure~\ref{FigMass} shows the spin-resolved differential decay rate $\mathrm{d}\Gamma/\mathrm{d}\varepsilon_+$ at fixed $\chi_a=1$, with the positron spin quantized along $\mathbf{B}$. The result in Fig.~\ref{FigMass}(a) is based on the same data as Fig.~\ref{FigdGamma}(b3), but is shown on a linear rather than logarithmic scale to provide the $\delta_m=0$ reference for comparison with the finite-mass spectra in Figs.~\ref{FigMass}(b)--\ref{FigMass}(d). As $\delta_m$ increases from zero toward the vacuum threshold, the finite ALP mass primarily enhances the overall yield while leaving the general spectral features largely unchanged. The antiparallel channel remains dominant throughout the full energy range. Above threshold, the opening of the vacuum decay channel further enhances the overall rate and introduces clear oscillatory modulations into the spectrum. These modulations are already visible in Fig.~\ref{FigMass}(c) and become increasingly pronounced as $\delta_m$ increases [Fig.~\ref{FigMass}(d)].

These features carry over to the integrated yield. Figure~\ref{FigMassTotal} shows the spin-resolved total rate, obtained by integrating the differential spectrum over $\varepsilon_+$, at fixed $\chi_a=1$ for three choices of the positron spin-quantization axis. As the mass approaches the threshold, the rate rises rapidly, and for $\delta_m>2$, oscillatory modulations develop in all spin channels. Because the positron spin is fixed in each panel, summing the parallel and antiparallel channels in Fig.~\ref{FigMassTotal} corresponds only to summing over the electron spin. For quantization along $\mathbf{T}$ or $\mathbf{N}$, no single-spin term survives this sum. For quantization along $\mathbf{B}$, however, the term proportional to $(\boldsymbol{\zeta}_{+}-\boldsymbol{\zeta}_{-})\cdot\mathbf{B}$ remains, so that the electron-spin-summed rate can still depend on the fixed positron polarization.

The different visibility of these oscillations among the three spin-quantization axes can be traced to the structure of the differential rate in Eq.~\eqref{eq:Airy}, which contains different combinations of the Airy-function contributions ${\rm Ai}'(u)/u$ and $\int_u^\infty {\rm Ai}(t)\,\mathrm{d}t$. Taking the parallel configuration as an example, the coefficients of these two terms appear in the ratios $2:1$, $3:1$, and $1:1$ for the $\mathbf{T}$, $\mathbf{N}$, and $\mathbf{B}$ directions, respectively. After integration over $\varepsilon_+$, these combinations undergo different degrees of cancellation, leading to different fringe contrasts for the three axes. For $\mathbf{T}$ and $\mathbf{N}$, the dominant oscillatory contribution is carried by the parallel spin configuration, producing the pronounced modulations of the black solid curves. Along $\mathbf{B}$, however, the additional spin-dependent term proportional to $(\boldsymbol{\zeta}_{+}-\boldsymbol{\zeta}_{-})\cdot\mathbf{B}$ redistributes the oscillation strength and shifts the dominant modulation to the antiparallel configuration.

The mass dependence of the dominant spin channel has a different origin. Even below threshold, where pair production is entirely field-induced, the two spin configurations evolve differently with \(\delta_m\). Their ordering may therefore reverse as the mass varies, leading to the crossings observed between the spin-resolved curves for the \(\mathbf T\) and \(\mathbf N\) axes in Figs.~\ref{FigMassTotal}(a) and~(b). Above threshold, the newly opened vacuum channel further modifies the relative populations of the two channels. Along \(\mathbf B\), by contrast, no crossing occurs, implying that the same spin channel remains dominant throughout the entire mass range, and the splitting between the two channels is the largest among the three axes, exceeding the moderate splitting along \(\mathbf T\) and the intermediate splitting along \(\mathbf N\).

\begin{figure}
\centering
\includegraphics[width=0.5\textwidth]{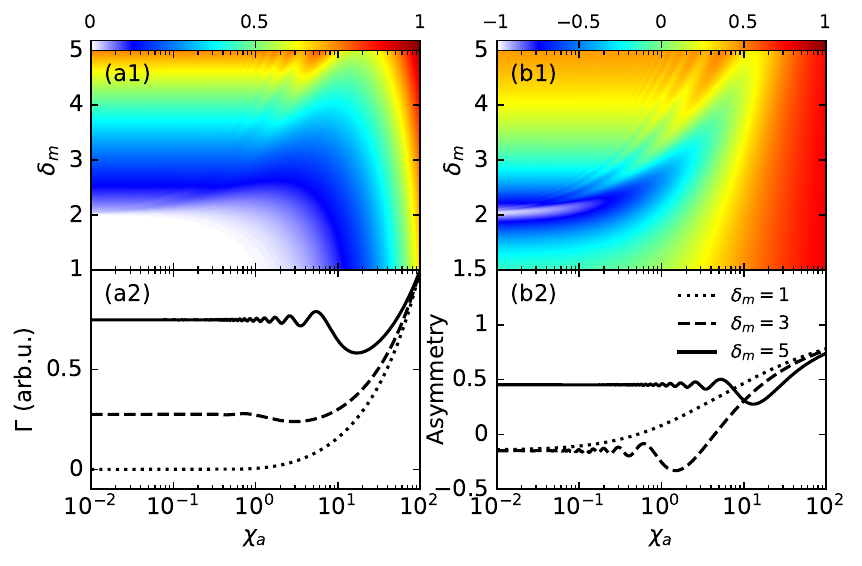}
\caption{%
Total pair-production rate and the corresponding spin asymmetry as functions of the quantum nonlinearity parameter \(\chi_a\) and the dimensionless ALP mass \(\delta_m\), for spin quantization along \(\mathbf{T}\). Panel~(a1): spin-summed total rate \(\Gamma\) in the \((\chi_a,\delta_m)\) plane. Panel~(b1): corresponding spin asymmetry \(\mathcal{A}\), defined in Eq.~\eqref{eq:asymmetry}. Panels (a2) and (b2) show representative cuts of panels (a1) and (b1) at \(\delta_m=1\), \(3\), and \(5\), represented by dotted, dashed, and solid curves, respectively. The ALP energy is fixed at \(\omega=1~\mathrm{GeV}\).}
\label{FigTotMass}
\end{figure}

To characterize these features more systematically, we focus on spin quantization along \(\mathbf{T}\) in Fig.~\ref{FigTotMass}. Panel~(a1) shows the spin-summed total rate in the \((\chi_a,\delta_m)\) plane, while panel~(b1) shows the corresponding spin asymmetry,
\begin{equation}
\label{eq:asymmetry}
\mathcal{A}=
\frac{\Gamma_{\uparrow_+\uparrow_-}-\Gamma_{\uparrow_+\downarrow_-}}
{\Gamma_{\uparrow_+\uparrow_-}+\Gamma_{\uparrow_+\downarrow_-}},
\end{equation}
where the arrows denote the positron and electron spin projections along the chosen quantization axis. Panels~(a2) and~(b2) show representative cuts for \(\delta_m=1\), \(3\), and \(5\).

Below threshold, the decay remains purely field-induced, and the total rate increases monotonically with \(\chi_a\), as illustrated by the \(\delta_m=1\) cut in Fig.~\ref{FigTotMass}(a2). Above threshold, interference fringes appear in both the two-dimensional map and the one-dimensional cuts. In particular, for \(\delta_m=3\) and \(5\), there is a clear interval over which the total rate decreases with increasing \(\chi_a\). Thus, a stronger laser background can suppress the decay, reminiscent of the suppression previously found for muon decay in strong electromagnetic fields~\cite{823w-2g4b}. The asymmetry \(\mathcal{A}\) exhibits the same threshold-induced reorganization. It varies smoothly with \(\chi_a\) for \(\delta_m<2\), but becomes strongly nonmonotonic above threshold and inherits the fringe pattern of the total rate [Figs.~\ref{FigTotMass}(b1) and~\ref{FigTotMass}(b2)].

\subsection{Estimate of interference extrema}

The oscillatory structures observed above threshold can be traced to the asymptotic form of the Airy function appearing in the rate. For a negative argument \(u<0\) with \(|u|\gg1\), the Airy function admits the expansion
\begin{equation}
\mathrm{Ai}(u) 
\approx 
\frac{1}{\sqrt{\pi}\,|u|^{1/4}}
\sin\!\left(
\frac{2}{3}|u|^{3/2} + \frac{\pi}{4}
\right).
\end{equation}
Using the previously defined expression \( u = \Delta\left( \frac{\omega^2}{\chi_a\,\varepsilon_-\varepsilon_+} \right)^{2/3}\), one finds that for \(\Delta<0\), the oscillatory phase behaves as
\begin{equation}
\Phi(\chi_a)\equiv\frac{2}{3}|u|^{3/2}=\frac{2}{3}\frac{|\Delta|^{3/2}\,\omega^2}{\chi_a \varepsilon_- \varepsilon_+}.
\end{equation}
The extrema of the oscillatory contribution are therefore expected to occur near the points satisfying
\(\Phi(\chi_a) \sim n\pi\), with \(n\in\mathbb N^+\). This condition leads to the parametric scaling
$\chi_a^{(n)}\sim\frac{2}{3\pi n}\,
\frac{|\Delta|^{3/2}\,\omega^2}{\varepsilon_- \varepsilon_+}$
for the locations of the interference extrema.

To estimate the fringe positions in the energy-integrated rate, we evaluate the phase at the symmetric energy partition, \(\varepsilon_+=\varepsilon_-=\omega/2\), for which \(\omega^2/(\varepsilon_-\varepsilon_+)=4\) and \(|\Delta|=\tfrac14(\delta_m^2-4)\). 
The scaling then reduces to
\begin{equation}
\label{eq:phasemateching2}
\chi_a^{(n)}
\sim
\frac{1}{3\pi n}\,(\delta_m^2-4)^{3/2}.
\end{equation}
The locations of the extrema therefore scale as \(1/n\), so the oscillations become progressively denser toward smaller \(\chi_a\).
Moreover, the characteristic field scale increases as \((\delta_m^2-4)^{3/2}\) with the ALP mass, consistent with the interference patterns displayed in Fig.~\ref{FigTotMass}. 
Although the precise locations of the extrema receive corrections from subleading terms and phase shifts, their leading parametric dependence is captured by this asymptotic estimate.

\subsection{Net polarization of produced positrons}
\label{sec:polarization}

We finally consider the net spin polarization acquired by the produced positron after summing over the electron spin and integrating over the energy partition. Although the parent ALP is spinless, the external electromagnetic field breaks the rotational symmetry of the vacuum decay and distinguishes different spin projections of the produced fermions. This allows the pair-production process to generate a nonzero net polarization.

\begin{figure}
\centering
\includegraphics[width=0.5\textwidth]{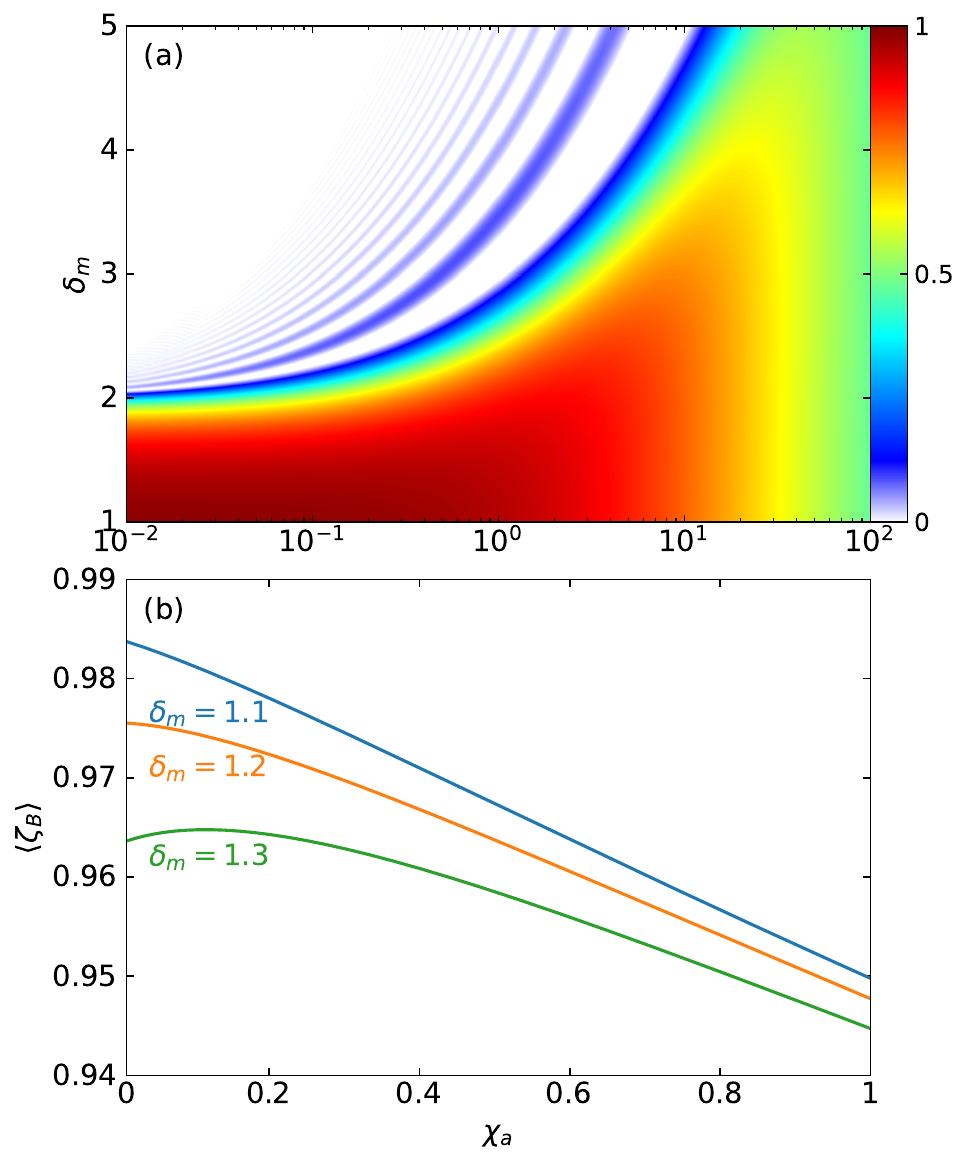}
\caption{%
Net positron polarization \(\langle\zeta_B\rangle\) [Eq.~\eqref{eq:pair_pol}] as a function of the quantum nonlinearity parameter~\(\chi_a\) and the dimensionless ALP mass \(\delta_m=m_a/m_e\), with the spin quantization axis along~\(\mathbf{B}\). Panel~(a) shows the polarization in the $(\chi_a,\delta_m)$ plane.
Panel~(b) shows representative below-threshold cuts for
$\delta_m=1.1$, $1.2$, and $1.3$, illustrating the crossover from
monotonic to nonmonotonic behavior near
$\delta_m^{(c)}\simeq1.2$.}
\label{FigNetF}
\end{figure}

By symmetry, the net positron polarization can be nonvanishing only along the \(\mathbf{B}\) direction, while the \(\mathbf{T}\) and \(\mathbf{N}\) components vanish identically. We therefore define
\begin{equation}
  \bigl\langle\zeta_B\bigr\rangle
  =\frac{\Gamma_{\uparrow_+\uparrow_-}
         +\Gamma_{\uparrow_+\downarrow_-}
         -\Gamma_{\downarrow_+\uparrow_-}
         -\Gamma_{\downarrow_+\downarrow_-}}
        {\Gamma_{\uparrow_+\uparrow_-}
         +\Gamma_{\uparrow_+\downarrow_-}
         +\Gamma_{\downarrow_+\uparrow_-}
         +\Gamma_{\downarrow_+\downarrow_-}}\,,
  \label{eq:pair_pol}
\end{equation}
where the arrows denote spin projections along~\(\mathbf{B}\). Figure~\ref{FigNetF} shows \(\langle\zeta_B\rangle\) in the \((\chi_a,\delta_m)\) plane, which is again organized around the threshold at \(\delta_m=2\).

In the weak-field, below-threshold regime, the energy integral is dominated by nearly symmetric sharing, $\varepsilon_+\simeq\varepsilon_-\simeq\omega/2$, which gives
$\left\langle \zeta_B \right\rangle_{\chi_a\to0}
\simeq
\frac{\sqrt{1-\delta_m^2/4}}
{1-\delta_m^2/8}$.
This approaches unity for $\delta_m\to0$ but vanishes as $\delta_m\to2^{-}$, showing that the weak-field polarization is strongly suppressed near threshold. The full numerical results reveal a crossover near $\delta_m^{(c)}\simeq1.2$. As shown in Fig.~\ref{FigNetF}(b), for $\delta_m<\delta_m^{(c)}$, the polarization decreases monotonically with increasing $\chi_a$. By contrast, for $\delta_m^{(c)}<\delta_m<2$, the polarization initially increases with $\chi_a$, reaches a maximum, and subsequently decreases toward the strong-field limit.
Above threshold, the net polarization drops and develops a strongly nonmonotonic fringe pattern, originating from the negative-argument oscillation of the Airy function in Eq.~\eqref{eq:Airy}. The fringe spacing decreases toward smaller \(\chi_a\), in qualitative agreement with the estimate in Eq.~\eqref{eq:phasemateching2}.

For sufficiently light ALPs, field-induced ALP decay below threshold can yield highly polarized positrons, while the mass- and field-dependent fringe pattern above threshold constitutes a distinctive spectroscopic signature of the Airy-function structure. The same spin-resolved rate also encodes nontrivial two-particle spin correlations, which we quantify through the entanglement analysis in the next section.

\section{Spin entanglement of the produced pair}
\label{sec:entanglement}

From the spin-resolved differential rate in Eq.~\eqref{eq:Airy}, we construct the reduced two-qubit spin density matrix of the produced \(e^+e^-\) pair and quantify the entanglement it encodes. We use the concurrence as a quantitative measure of entanglement and the spin correlator \(c_\Sigma=\langle\boldsymbol{\sigma}_+\!\cdot\!\boldsymbol{\sigma}_-\rangle\) to distinguish singlet-like from triplet-like correlations. Since Eq.~\eqref{eq:Airy} is derived for a CCF, the results below should be understood within the same approximation.

\subsection{Reduced density matrix}
\label{subsec:density_matrix}

For a final state with fixed positron energy fraction \(\delta_+\), the two-particle spin Hilbert space is
\(\mathcal{H}=\mathcal{H}_+\otimes\mathcal{H}_-\),
where \(\mathcal{H}_+\) and \(\mathcal{H}_-\) denote the positron and electron spin spaces, respectively.
Using the instantaneous orthonormal triad
\(\{\mathbf{T},\mathbf{N},\mathbf{B}\}\) introduced in Sec.~\ref{sec:FS}, we define the projected Pauli operators
\(\sigma_T\equiv \boldsymbol{\sigma}\!\cdot\!\mathbf{T}\),
\(\sigma_N\equiv \boldsymbol{\sigma}\!\cdot\!\mathbf{N}\), and
\(\sigma_B\equiv \boldsymbol{\sigma}\!\cdot\!\mathbf{B}\).
After the angular degrees of freedom have been traced out, the reduced spin density matrix conditioned on the fixed energy partition \(\delta_+\) can be written as
\begin{equation}
\begin{split}
\rho=\frac{1}{4}\Bigl[&\,\mathbb{I}\otimes\mathbb{I}
+c_p\bigl(\sigma_B\otimes\mathbb{I}-\mathbb{I}\otimes\sigma_B\bigr)\\
&+c_T\,\sigma_T\otimes\sigma_T
+c_N\,\sigma_N\otimes\sigma_N
+c_B\,\sigma_B\otimes\sigma_B\Bigr],
\end{split}
\label{eq:rho_tnb}
\end{equation}
where
\begin{equation}
\begin{aligned}
c_p&=\sqrt{\frac{\Delta}{u}}\frac{\mathrm{Ai}(u)}{C_0},\\[4pt]
c_T&=-1-\frac{\Delta}{C_0}\left[2\,\frac{\mathrm{Ai}'(u)}{u}+\int_u^\infty \mathrm{Ai}(t)\,\mathrm{d}t\right],\\[4pt]
c_N&=-1-\frac{\Delta}{2C_0}\left[3\,\frac{\mathrm{Ai}'(u)}{u}+\int_u^\infty \mathrm{Ai}(t)\,\mathrm{d}t\right],\\[4pt]
c_B&=-1-\frac{\Delta}{2C_0}\left[\frac{\mathrm{Ai}'(u)}{u}+\int_u^\infty \mathrm{Ai}(t)\,\mathrm{d}t\right],
\end{aligned}
\label{eq:Cs}
\end{equation}
and
\begin{equation}
C_0=-\Delta \frac{\mathrm{Ai}'(u)}{u}
+\frac{1}{2}\,\frac{m_a^2}{m_e^2}\,
\frac{\varepsilon_+}{\omega}\,\frac{\varepsilon_-}{\omega}
\int_u^\infty \mathrm{Ai}(t)\,\mathrm{d}t.
\label{eq:C0_def}
\end{equation}
We have numerically verified that the constructed \(\rho\) is Hermitian, normalized, and positive semidefinite for all investigated parameters.

With $\mathbf{B}$ chosen as the spin-quantization axis and \(\mathbf{T}\leftrightarrow x\), \(\mathbf{N}\leftrightarrow y\), and \(\mathbf{B}\leftrightarrow z\), the density matrix takes the explicit form
\begin{equation}
\rho=\frac{1}{4}
\begin{pmatrix}
1+c_B & 0 & 0 & c_T-c_N \\[2pt]
0 & 1-c_B+2c_p & c_T+c_N & 0 \\[2pt]
0 & c_T+c_N & 1-c_B-2c_p & 0 \\[2pt]
c_T-c_N & 0 & 0 & 1+c_B
\end{pmatrix}
\label{eq:rho_Xstate}
\end{equation}
in the ordered basis
\(\{\ket{\uparrow_+\uparrow_-},\ket{\uparrow_+\downarrow_-},\ket{\downarrow_+\uparrow_-},\ket{\downarrow_+\downarrow_-}\}_{\mathbf B}\).
This is a standard two-qubit \(X\) state, for which the concurrence can be obtained in closed form.
The concurrence is
\begin{equation}
\mathcal{C}=\max\!\bigl\{0,\mathcal{C}_1,\mathcal{C}_2\bigr\},
\label{eq:concurrence_main}
\end{equation}
with
\begin{equation}
\begin{aligned}
\mathcal{C}_1&=\frac{1}{2}\Bigl[|c_T+c_N|-(1+c_B)\Bigr],\\
\mathcal{C}_2&=\frac{1}{2}\Bigl[|c_T-c_N|-\sqrt{(1-c_B)^2-4c_p^2}\Bigr].
\end{aligned}
\end{equation}
For two-qubit states, \(\mathcal{C}>0\) is equivalent to entanglement and hence to a partial transpose with a negative eigenvalue. 
The positive-partial-transpose criterion is necessary and sufficient for separability in \(2\times2\) systems \cite{peres1996separability,horodecki2001separability}.
In what follows, \(\mathcal{C}=0\) corresponds to a separable state, \(0<\mathcal{C}<1\) to a partially entangled state, and \(\mathcal{C}=1\) to a maximally entangled state.

Finally, the spin correlator introduced above evaluates to
\[
c_\Sigma=\langle \boldsymbol{\sigma}_+\!\cdot\!\boldsymbol{\sigma}_-\rangle=c_T+c_N+c_B,
\]
and determines the expectation value of the total spin squared,
\begin{equation}
\langle S^2\rangle=\frac{3}{2}+\frac{1}{2}c_\Sigma.
\label{eq:S2_expect}
\end{equation}
A pure singlet state corresponds to \(c_\Sigma=-3\), whereas a pure triplet state gives \(c_\Sigma=+1\). Thus, negative values of \(c_\Sigma\) indicate singlet-like correlations, while positive values indicate triplet-like correlations.

\subsection{Analytical limits and entanglement mechanisms}
\label{subsec:analytical_limits}

For physical interpretation, it is convenient to introduce the singlet--triplet basis
\begin{align}
\ket{S}&=\frac{1}{\sqrt{2}}\bigl(\ket{\uparrow\downarrow}-\ket{\downarrow\uparrow}\bigr),\\[4pt]
\ket{T_0}&=\frac{1}{\sqrt{2}}\bigl(\ket{\uparrow\downarrow}+\ket{\downarrow\uparrow}\bigr),\\[4pt]
\ket{T_+}&=\ket{\uparrow\uparrow},\qquad
\ket{T_-}=\ket{\downarrow\downarrow}.
\end{align}

\paragraph{Strong-field limit: field-induced, triplet-like entanglement.}
In the strong-field limit, one has \(u\to0\), and therefore
\begin{equation}
c_T\to 1,\qquad
c_N\to \frac{1}{2},\qquad
c_B\to -\frac{1}{2},\qquad
c_p\to 0.
\label{eq:strong_lim}
\end{equation}
Accordingly,
\begin{equation}
c_\Sigma\to 1,\qquad
\langle S^2\rangle\to 2,
\end{equation}
showing that the reduced spin state is entirely supported in the triplet sector. In this limit,
\begin{equation}
\rho\to \frac{3}{4}\ket{T_0}\!\bra{T_0}+\frac{1}{4}\ket{\phi_T}\!\bra{\phi_T},
\qquad
\ket{\phi_T}\equiv \frac{\ket{T_+}+\ket{T_-}}{\sqrt{2}},
\label{eq:rho_strong}
\end{equation}
which represents an anisotropic mixture within the triplet manifold induced by the preferred direction of the background field. The concurrence approaches
\begin{equation}
\mathcal{C}\to \frac{1}{2}.
\label{eq:rho_strongC}
\end{equation}

\paragraph{Weak-field, above-threshold regime: singlet-to-triplet crossover.}
We next consider the vacuum-dominated above-threshold regime, \(\chi_a\ll1\) and \(\delta_m>2\). It is convenient to introduce the kinematic variable
\begin{equation}
\mathcal{X}\equiv\frac{m_a^2}{m_e^2}\,\delta_+(1-\delta_+)=1-\Delta.
\label{eq:largeX}
\end{equation}
In vacuum, the decay is kinematically allowed for \(\mathcal{X}\geq1\), with \(\mathcal{X}=1\) defining the boundary of the allowed energy interval. This boundary is reached either as \(m_a\to2m_e\) with symmetric energy sharing or, for fixed \(m_a>2m_e\), as \(\delta_+\to\delta_+^{\min,\max}\). In the presence of the background field, configurations with \(\mathcal{X}<1\) are also accessible.

Using the vacuum expression in Eq.~\eqref{eq:vac_differential}, the spin-correlation coefficients approach
\begin{equation}
c_T\to 1-\frac{2}{\mathcal{X}},\qquad
c_N\to-\frac{1}{\mathcal{X}},\qquad
c_B\to-\frac{1}{\mathcal{X}},\qquad
c_p\to0,
\label{eq:vac_crossover_lim}
\end{equation}
so that
\begin{equation}
c_\Sigma\to 1-\frac{4}{\mathcal{X}},\qquad
\langle S^2\rangle\to 2-\frac{2}{\mathcal{X}}.
\label{eq:cSigma_crossover}
\end{equation}
In this regime, the reduced spin state is governed entirely by the single kinematic variable \(\mathcal{X}\), evolving continuously from singlet-like to triplet-like character as \(\mathcal{X}\) increases.

Within the vacuum-dominated weak-field limit considered here, taking \(\mathcal{X}\to1\) gives
\begin{equation}
c_T,\,c_N,\,c_B\to-1,\qquad c_\Sigma\to-3,
\end{equation}
and the reduced density matrix approaches the pure spin-singlet state,
\begin{equation}
\rho\to\ket{S}\!\bra{S},\qquad \mathcal{C}\to1.
\label{eq:rho_singlet}
\end{equation}
This is precisely the familiar \(^{1}S_{0}\) configuration of the vacuum decay \(a\to e^+e^-\). 
In the ALP rest frame, conservation of angular momentum and parity for a pseudoscalar $J^P=0^-$ state selects the spin-singlet $S$-wave channel. The resulting electron--positron pair is therefore maximally spin entangled at fixed momentum.

In the opposite limit, \(\mathcal{X}\gg1\), corresponding to large \(m_a\) away from the extreme energy-partition boundaries, one finds
\begin{equation}
c_T\to1,\qquad c_N,\,c_B\to0,\qquad c_\Sigma\to1,
\end{equation}
and
\begin{equation}
\rho\to\frac12\ket{T_0}\!\bra{T_0}
+\frac12\ket{\phi_T}\!\bra{\phi_T},
\qquad \mathcal{C}\to0.
\label{eq:rho_weak_largemass}
\end{equation}
Thus, the reduced spin state becomes triplet-like but separable.


The contrast between the near-threshold singlet state in Eq.~\eqref{eq:rho_singlet} and the large-mass separable mixture in Eq.~\eqref{eq:rho_weak_largemass} originates from tracing over the momentum degrees of freedom. Since the spin state depends on the emission momentum in the relativistic basis used here, the transverse integration over $(\alpha,\beta)$ combines different momentum-conditioned spin states into a mixed reduced state, reducing the concurrence.

The degree of mixing is therefore controlled both by the range of momentum configurations contributing at fixed \(\delta_+\) and by the variation of the corresponding spin states across that range. In the limit \(\mathcal{X}\to1\), the available momentum range collapses. As \(m_a\to2m_e\) with \(\delta_+\to1/2\), the relative momentum of the pair in the ALP rest frame vanishes. For fixed \(m_a>2m_e\), \(\mathcal{X}\to1\) is instead reached at the kinematic endpoints \(\delta_+\to\delta_+^{\min,\max}\), where the allowed angular phase space collapses to a collinear configuration. In both cases, the momentum-induced mixing is suppressed, and the reduced spin state approaches the pure singlet limit of Eq.~\eqref{eq:rho_singlet}. Although the formation region grows as \(\mathcal{X}\to1\), the LCFA ordering can still be maintained parametrically in the adiabatic limit \(\omega_L\to0\); see also the discussion in Sec.~\ref{sec:vacuum}.

\subsection{Numerical results}
\label{subsec:numerical_results}

\begin{figure}[t]
\centering
\includegraphics[width=\columnwidth]{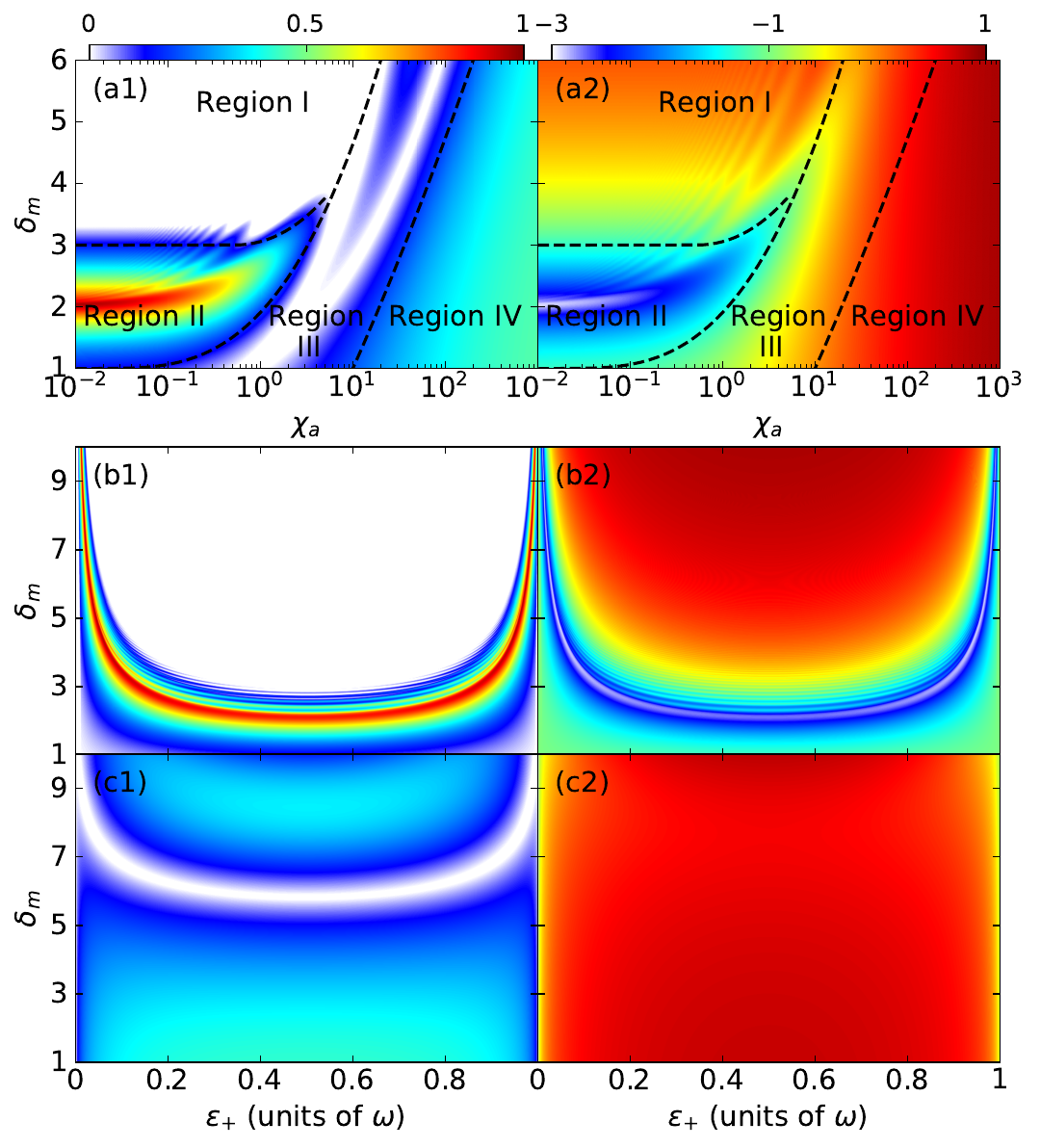}
\caption{Spin-entanglement properties of the produced \(e^+e^-\) pair. Panels (a1) and (a2) show the concurrence \(\mathcal{C}\) and the spin--spin correlator \(c_{\Sigma}=\langle\boldsymbol{\sigma}_{+}\cdot\boldsymbol{\sigma}_{-}\rangle\), respectively, evaluated from the energy-integrated reduced density matrix defined by Eq.~\eqref{eq:averaged_coefficients} in the \((\chi_a,\delta_m)\) parameter space. The dashed curves serve as guides to the eye separating regions with distinct spin-entanglement properties. Regions I--IV correspond to a triplet-like separable region, a singlet-like entangled region, an intermediate region, and a triplet-like entangled region, respectively.
The corresponding differential structures in the \((\delta_{+},\delta_m)\) plane are shown in panels (b1) and (b2) for \(\chi_a=0.1\), and in panels (c1) and (c2) for \(\chi_a=100\).}
\label{fig:entanglement}
\end{figure}

We now map the entanglement structure numerically across the full \((\chi_a,\delta_m)\) parameter space. To characterize the overall parameter dependence, we first consider the reduced density matrix integrated over the energy partition. Since the spin-summed differential rate is proportional to \(C_0/[\delta_+(1-\delta_+)]\), with \(C_0\) defined in Eq.~\eqref{eq:C0_def}, we average the coefficients in Eq.~\eqref{eq:Cs} over the positron energy fraction according to
\begin{equation}
\bar{c}_i=\frac{\int_0^1\mathrm{d}\delta_+\,C_0\,c_i/[\delta_+(1-\delta_+)]}{\int_0^1\mathrm{d}\delta_+\,C_0/[\delta_+(1-\delta_+)]}\,,\qquad i\in\{p,T,N,B\}\,.
\label{eq:averaged_coefficients}
\end{equation}
The averaged coefficients \(\bar{c}_i\) then replace \(c_i\) in Eq.~\eqref{eq:rho_tnb} to define the energy-integrated spin density matrix. Physically, this corresponds to a measurement in which the energy partition of the pair is not resolved. The concurrence \(\mathcal{C}\) and the spin--spin correlator \(c_\Sigma\) obtained from this density matrix are shown in Figs.~\ref{fig:entanglement}(a1) and (a2).

In the weak-field limit, the spin-entanglement structure is highly sensitive to the opening of the vacuum decay channel because field-induced pair production is tunneling suppressed. At \(\delta_m=0\), Eq.~\eqref{eq:weak_ALP} shows that pair production is dominated by the configuration in which the positron spin is parallel to \(\mathbf{B}\) and the electron spin is antiparallel to \(\mathbf{B}\). This strong preference for a single product-spin configuration leads to a nearly vanishing concurrence.
As $\delta_m$ increases from zero toward the threshold, the concurrence grows, and the produced pair becomes nearly singlet-like and strongly entangled, in agreement with the analytical limit in Eq.~\eqref{eq:rho_singlet}. As $\delta_m$ increases further beyond the threshold, the overall trend reverses, with the concurrence decreasing while $c_\Sigma$ evolves toward a triplet-like value and approaches the separable vacuum-dominated limit of Eq.~\eqref{eq:rho_weak_largemass}. Superimposed on this smooth trend, pronounced fringes appear in both observables once the vacuum channel becomes kinematically open. These fringes are the entanglement counterpart of the vacuum--field interference oscillations in the rate discussed in Sec.~\ref{sec:above}.

In contrast, in the strong-field limit, the pair-production dynamics is dominated by the external field over a broad range of ALP masses. Accordingly, both \(\mathcal{C}\) and \(c_\Sigma\) depend only weakly on \(\delta_m\), and the threshold-related structures associated with the vacuum channel are largely suppressed. The spin--spin correlator remains predominantly triplet-like, while the concurrence stays finite, consistent with the field-induced limit \(\mathcal{C}\to1/2\) derived in Eq.~\eqref{eq:rho_strongC}.

To identify the energy-resolved structures underlying this integrated behavior, Figs.~\ref{fig:entanglement}(b1)--(c2) show the differential concurrence \(\mathcal{C}\) and \(c_\Sigma\) in the \((\delta_+,\delta_m)\) plane for two representative field strengths. For \(\chi_a=0.1\) [Figs.~\ref{fig:entanglement}(b1) and (b2)], the pattern is well described by the vacuum expression in Eq.~\eqref{eq:vac_crossover_lim} and is therefore governed mainly by the single kinematic variable \(\mathcal{X}\) introduced in Eq.~\eqref{eq:largeX}.
Near the boundary \(\mathcal X=1\), the produced pair becomes strongly entangled and increasingly singlet-like, consistent with the vacuum-dominated limit discussed in Sec.~\ref{subsec:analytical_limits}.
Deep inside the vacuum-allowed region, where \(\mathcal{X}\gg1\), the spin state instead evolves toward a separable triplet-like mixture. Below threshold, \(\delta_m<2\), the process is purely field-induced, and both observables vary smoothly with \(\delta_+\) and \(\delta_m\). In the crossover region between these two descriptions, oscillatory structures develop, mirroring the fringes seen in the integrated observables.

The strong-field cut at \(\chi_a=100\) [Figs.~\ref{fig:entanglement}(c1) and (c2)] displays a qualitatively different pattern that is well captured by the field-induced strong-field limit. Over most of the parameter region, \(c_\Sigma\) remains close to the triplet-like value, while the concurrence stays finite and approaches \(\mathcal{C}=1/2\). Only weak modulations persist at larger \(\delta_m\), reflecting the residual influence of the vacuum contribution. In this strongly field-dominated regime, the produced electron and positron generally continue to interact with the background field and may undergo subsequent strong-field QED processes, such as photon emission or cascade formation. These processes can further modify the spin correlations before detection. A quantitative description of the subsequent dynamics requires going beyond the present single-vertex treatment and lies outside the scope of this work.

\section{Conclusion}
\label{sec:conclusion}

We have derived the spin-resolved differential decay rate for a relativistic ALP decaying into an electron--positron pair in a CCF using the Baier--Katkov quasiclassical operator method. The final result, presented in the unified Airy-function form of Eq.~\eqref{eq:Airy}, provides a compact analytical expression that consistently incorporates both the finite ALP mass and the spin degrees of freedom of the produced pair, while smoothly connecting the vacuum-forbidden and vacuum-allowed kinematic regions across \(\Delta=0\).

For $m_a<2m_e$, pair production is entirely field-induced. In the massless limit, the decay rate exhibits exponential suppression through the factor $\exp[-8/(3\chi_a)]$ in weak fields and follows the $\chi_a^{2/3}$ scaling in strong fields, matching the corresponding asymptotic scalings of photon-induced pair creation. At the same time, the pseudoscalar coupling \(\bar{\psi}\gamma^5\psi\) leads to qualitatively distinct spin-resolved channels and spin-correlation patterns in the produced pair.
For \(m_a>2m_e\), the vacuum decay channel opens over a finite energy interval and contributes to the same final state as the field-assisted process. 
Their coherent interplay, reflected in the two-real-saddle structure, produces oscillatory modulations in the energy spectrum, the total rate, and the spin asymmetry.

By constructing the reduced two-particle spin density matrix from Eq.~\eqref{eq:Airy}, we have identified qualitatively distinct correlation patterns across the \((\chi_a,\delta_m)\) plane. The produced spin state is triplet-like with finite entanglement in the strong-field regime, triplet-like but separable deep in the vacuum-dominated regime, and nearly singlet-like with strong entanglement in the vacuum-threshold crossover region. Thus, the quantum state of the produced pair carries a direct imprint of the underlying decay mechanism.
These results indicate that the concurrence and spin correlations of the produced pair may provide complementary, process-specific information on ALP-mediated pair production. In particular, such observables encode aspects of the underlying pseudoscalar coupling that are not fully captured by spin-summed rates alone. This perspective is in the same spirit as recent proposals and measurements using entanglement and spin correlations as probes of new physics in high-energy collisions~\cite{atlas2024observation,cms2024observation,ruzi2024testing,barr2024quantum}.

Finally, the framework developed here provides both a spin-resolved description of ALP decay in strong electromagnetic fields and a practical means of tracking the quantum state of the produced electron--positron pair. Its compact analytical form makes it well suited for implementation in semiclassical Monte Carlo simulations of strong-field particle cascades, in which ALP production and decay can be incorporated on the same footing as standard QED processes. 
A quantitative assessment of experimental sensitivity would require an end-to-end simulation including a specified relativistic ALP source, realistic laser geometry and overlap, QED backgrounds, detector acceptance and efficiency, and the subsequent dynamics and spin analysis of the produced pair. Such an experimental sensitivity study lies beyond the scope of the present work.

\vspace{6pt}
\textit{Acknowledgments}\textemdash
This work is supported by the National Natural Science Foundation of China (Grants No. 12574377 and No. 12474312) and the National Key R\&D Program of China (Grant No. 2021YFA1601700).
The computations were performed on the Siyuan-1 cluster supported by the Center for High Performance Computing at Shanghai Jiao Tong University. 
P.-L. H. acknowledges support from the Pujiang Program of the Shanghai Baiyulan Talent Plan (Grant No.~24PJA046), the Xiaomi Young Scholar Program, the Shanghai Jiao Tong University 2030 Initiative, and the Yangyang Development Fund.
X. M. acknowledges support from the Postdoctoral Fellowship Program of CPSF under Grant No.~GZB20260818, the China Postdoctoral Science Foundation under Grant No.~2026M793723, and the Shanghai Super Postdoctoral Fellowship Program.

\appendix
\section{Crossing symmetry with spin}
\label{app:crossing_spin}

In this appendix, we briefly summarize how the spin label transforms under crossing symmetry. Since the crossing transformation of the on-shell four-momentum is standard, we do not repeat it here and focus only on the corresponding mapping of the spin polarization.

Following the convention of Ref.~\cite{he2025semiclassical}, we write the positive-energy bispinor as
\begin{equation}
U(p,\boldsymbol{\zeta}_-)=\frac{1}{\sqrt{2\varepsilon}}\begin{pmatrix}
\sqrt{\varepsilon+m}\, w \\
\dfrac{\mathbf p\!\cdot\!\boldsymbol{\sigma}}{\sqrt{\varepsilon+m}}\, w
\end{pmatrix},
\qquad
p^\mu=(\varepsilon,\mathbf p),
\label{eq:U_spinor_crossing}
\end{equation}
where \(w^\dagger w=1\) and the corresponding spin-polarization vector is
\begin{equation}
\boldsymbol{\zeta}_- = w^\dagger \boldsymbol{\sigma} w.
\end{equation}

Under crossing, one replaces $p\to -p$. Defining
\begin{equation}
w'=\hat{\mathbf p}\cdot\boldsymbol{\sigma}\,w,
\qquad
\hat{\mathbf p}\equiv\frac{\mathbf p}{|\mathbf p|},
\end{equation}
the spinor with crossed momentum can be rewritten as
\begin{equation}
U(-p,\boldsymbol{\zeta}_{-})
=
\frac{1}{\sqrt{2\varepsilon}}
\begin{pmatrix}
\dfrac{\mathbf p\cdot\boldsymbol{\sigma}}{\sqrt{\varepsilon+m}}\,w'\\
\sqrt{\varepsilon+m}\,w'
\end{pmatrix},
\label{eq:Uminus_rewrite_crossing}
\end{equation}
which coincides with the standard form of the negative-energy spinor
$V(p,\boldsymbol{\zeta}_{+})$, up to an irrelevant overall phase.
The Pauli polarization of the two-component spinor $w'$ is opposite to the physical positron rest-frame polarization, so that
\begin{equation}
\boldsymbol{\zeta}_{+}
=
-w'^\dagger\boldsymbol{\sigma}w'
=
\boldsymbol{\zeta}_{-}
-2(\hat{\mathbf p}\cdot\boldsymbol{\zeta}_{-})\hat{\mathbf p}.
\label{eq:zeta_reflection_crossing}
\end{equation}
This transformation is involutive: it leaves the component transverse to $\hat{\mathbf p}$ unchanged while reversing the longitudinal component.

Collecting the above results, and adopting the phase convention for the negative-energy spinor used here, we obtain the spin-resolved crossing relation
\begin{equation}
U(-p,\boldsymbol{\zeta}_{-})
=
V\!\left(p,\boldsymbol{\zeta}_{+}\right).
\label{eq:crossing_spin_final}
\end{equation}

\section{Airy--Bessel relations in the oscillatory regime}
\label{app:Bessel}

\subsection*{\texorpdfstring{Bessel combinations $\mathcal{J}_\pm$}{Bessel combinations}}
\label{app:Jpm}

In the oscillatory regime, \(b<0\) [see Appendix~\ref{app:identities}], the \(\tau\)-integration gives rise to the combinations
\begin{equation}
  \mathcal{J}_{\pm}(x)
  =
  \begin{cases}
    J_{1/3}(x) + J_{-1/3}(x), & (+),\\[6pt]
    J_{2/3}(x) - J_{-2/3}(x), & (-),
  \end{cases}
  \label{eq:Jpm}
\end{equation}
which repeatedly appear in the evaluation of the pair-production rate in the region \(\Delta<0\).

\subsection*{Airy--Bessel identities}
\label{app:Airy}

The combinations in Eq.~\eqref{eq:Jpm} are directly related to the Airy function and its derivative. Defining
\begin{equation}
  \zeta=\frac{2}{3}|x|^{3/2},
\end{equation}
one finds, for \(x>0\),
\begin{align}
  \mathrm{Ai}(x)
    &= \frac{1}{\pi}\sqrt{\frac{x}{3}}\,
       K_{1/3}(\zeta),
  \label{eq:Ai_pos}\\[4pt]
  \mathrm{Ai}'(x)
    &= -\frac{x}{\sqrt{3}\,\pi}\,
       K_{2/3}(\zeta),
  \label{eq:Aip_pos}
\end{align}
where \(K_\nu\) is the modified Bessel function of the second kind.
For \(x<0\), one has
\begin{align}
  \mathrm{Ai}(x)
    &= \frac{1}{3}\sqrt{-x}\,
       \mathcal{J}_+(\zeta),
  \label{eq:Ai_neg}\\[4pt]
  \mathrm{Ai}'(x)
    &= \frac{-x}{3}\,
       \mathcal{J}_-(\zeta).
  \label{eq:Aip_neg}
\end{align}
These identities make explicit the equivalence between the Airy-function representation in Eq.~\eqref{eq:Airy} and the Bessel-function forms in Eqs.~\eqref{eq:below} and~\eqref{eq:above}.

\section{Auxiliary integration identities}
\label{app:identities}

This appendix collects the elementary integrals needed to
evaluate the master integral in Eq.~\eqref{eq:Inttable}.
They are grouped by the variable of integration.

\subsection*{Time integration}
\label{app:tau}

The $\tau$~integrals take different forms depending on the sign of the coefficient~$b$ in the exponent $e^{-i(a\tau^3+b\tau)}$ (with $a>0$ throughout).

\subsubsection*{\texorpdfstring{Case~I: $b>0$
  (modified Bessel functions)}{Case I: b>0
  (modified Bessel functions)}}

\begin{equation}
\begin{aligned}
  \int_{-\infty}^{\infty}\!\mathrm{d}\tau\;
    e^{-i(a\tau^3+b\tau)}
  &= \frac{2}{\sqrt{3}}\,
     \sqrt{\frac{b}{3a}}\;
     K_{1/3}(z_\rho),
  \\[6pt]
  \int_{-\infty}^{\infty}\!\mathrm{d}\tau\;
    \tau\, e^{-i(a\tau^3+b\tau)}
  &= -\frac{2i}{\sqrt{3}}\,
     \frac{b}{3a}\;
     K_{2/3}(z_\rho),
  \\[6pt]
  \int_{-\infty}^{\infty}\!\mathrm{d}\tau\;
    \tau^2\, e^{-i(a\tau^3+b\tau)}
  &= -\frac{2}{\sqrt{3}}\,
     \biggl(\frac{b}{3a}\biggr)^{\!3/2}
     K_{1/3}(z_\rho),
\end{aligned}
\label{eq:intK}
\end{equation}
where $z_\rho = \frac{2}{3\sqrt{3}}\,b^{3/2}/\sqrt{a}$.

\subsubsection*{\texorpdfstring{Case~II: $b<0$
  (ordinary Bessel functions)}{Case II: b<0
  (ordinary Bessel functions)}}

\begin{equation}
\begin{aligned}
  \int_{-\infty}^{\infty}\!\mathrm{d}\tau\;
    e^{-i(a\tau^3+b\tau)}
  &= \frac{2\pi}{3}\,
     \sqrt{\frac{-b}{3a}}\;
     \mathcal{J}_+(z_\rho),
  \\[6pt]
  \int_{-\infty}^{\infty}\!\mathrm{d}\tau\;
    \tau\, e^{-i(a\tau^3+b\tau)}
  &= \frac{2\pi i}{3}\,
     \frac{-b}{3a}\;
     \mathcal{J}_-(z_\rho),
  \\[6pt]
  \int_{-\infty}^{\infty}\!\mathrm{d}\tau\;
    \tau^2\, e^{-i(a\tau^3+b\tau)}
  &= \frac{2\pi}{3}\,
     \biggl(\frac{-b}{3a}\biggr)^{\!3/2}
     \mathcal{J}_+(z_\rho),
\end{aligned}
\label{eq:intB}
\end{equation}
with $z_\rho = \frac{2}{3\sqrt{3}}\,(-b)^{3/2}/\sqrt{a}$ and $\mathcal{J}_\pm$ defined in Eq.~\eqref{eq:Jpm}.

\subsection*{Radial integration}
\label{app:radial}

After the azimuthal integration is performed (see Appendix~\ref{app:integral}), the remaining radial integrals fall into two classes according to the sign of~$b$.

\subsubsection*{\texorpdfstring{Region $b>0$
  (modified Bessel functions)}{Region b>0
  (modified Bessel functions)}}

\begin{equation}
\begin{aligned}
  \int_0^{\infty}\!\mathrm{d}x\;
    x^{1/3}\, K_{2/3}(x\,z)
  &= \frac{\Gamma(1/3)}{2^{2/3}\, z^{4/3}},
  \\[6pt]
  \int_0^{\infty}\!\mathrm{d}x\;
    x^{2/3}\, K_{1/3}(x\,z)
  &= \frac{\Gamma(2/3)}{2^{1/3}\, z^{5/3}},
  \\[6pt]
  \int_0^{\infty}\!\mathrm{d}x\;
    K_{1/3}(x\,z)
  &= \frac{\pi}{\sqrt{3}\, z}.
\end{aligned}
\label{eq:intK2}
\end{equation}

\subsubsection*{\texorpdfstring{Region $b<0$
  (ordinary Bessel functions)}{Region b<0
  (ordinary Bessel functions)}}

\begin{equation}
\begin{aligned}
  \int_0^1\! x^{1/3}\,
    \mathcal{J}_-(z\,x)\,\mathrm{d}x
  &= -\frac{\mathcal{J}_+(z)}{z}
     +\frac{2^{1/3}}{z^{4/3}\,\Gamma(2/3)},
  \\[8pt]
  \int_0^1\! x^{2/3}\,
    \mathcal{J}_+(z\,x)\,\mathrm{d}x
  &= \frac{\mathcal{J}_-(z)}{z}
     +\frac{2^{2/3}}{z^{5/3}\,\Gamma(1/3)},
  \\[8pt]
  \int_0^1\!
    \mathcal{J}_+(z\,x)\,\mathrm{d}x
  &= -\frac{1}{z}
     \int_{z}^{\infty}\!
       \mathcal{J}_+(x)\,\mathrm{d}x
     +\frac{2}{z}.
\end{aligned}
\label{eq:intB2}
\end{equation}

\section{\texorpdfstring{Integration formulas for $\Delta<0$}%
         {Integration formulas for Delta<0}}
\label{app:integral}

Combining the identities of Appendices~\ref{app:Jpm} and~\ref{app:identities}, we now evaluate the master integral appearing in Eq.~\eqref{eq:DP} for the case $\Delta<0$:
\begin{equation}
  F
  = \int_{-\infty}^{\infty}\!\mathrm{d}\alpha
    \int_{-\infty}^{\infty}\!\mathrm{d}\beta
    \int_{-\infty}^{\infty}\!\mathrm{d}\tau\;
    f(\tau,\alpha,\beta)\;
    e^{-i(a\tau^3+b\tau)},
  \label{eq:Inttable}
\end{equation}
with the coefficients $a$ and $b$ given in Eq.~\eqref{eq:ab}.

\paragraph{Angular integration.}
Introducing polar coordinates $(\rho,\varphi)$ in the $(\alpha,\beta)$~plane, where $\rho=\sqrt{\alpha^2+\beta^2}$, the azimuthal integration is elementary: isotropic terms
acquire a factor of~$2\pi$, the $\alpha^2$ and $\beta^2$ terms each yield a factor of~$\pi$, and the cross term~$\alpha\beta$ vanishes.

\paragraph{Radial integration.}
For $\Delta<0$, the coefficient~$b$ changes sign at the radius
\begin{equation}
  \rho_0
  = \frac{m_e}{\varepsilon_+}\sqrt{-\Delta}\,,
\end{equation}
which divides the radial domain into two regions:
\begin{enumerate}
\item $\rho<\rho_0$ ($b<0$):
  the $\tau$~integral is evaluated using
  Eq.~\eqref{eq:intB}; the subsequent radial integral is
  performed with Eq.~\eqref{eq:intB2}.
\item $\rho>\rho_0$ ($b>0$):
  the $\tau$~integral yields modified Bessel functions via
  Eq.~\eqref{eq:intK}; the radial integral is then evaluated
  with Eq.~\eqref{eq:intK2}.
\end{enumerate}
Summing the contributions from both regions gives the results
collected in Table~\ref{TableInt}, expressed in terms of the
combinations $\mathcal{J}_\pm$ defined in Eq.~\eqref{eq:Jpm}.

\begin{table}
\centering
\caption{%
  Results for the integral~\eqref{eq:Inttable} when $\Delta<0$.
  The argument of the Bessel-function combinations is
  $z = \frac{2}{3\chi_a}\,
       \frac{\omega^2}{\varepsilon_-\varepsilon_+}\,
       (-\Delta)^{3/2}$.}
\label{TableInt}
\begin{tabular}{c|l}
\toprule
$f$ & $F$ \\
\midrule
$1$ &
$\displaystyle
  \frac{4\pi^2}{3}\,
  \frac{\varepsilon_-}{\varepsilon_+}\,
  \frac{1}{\omega}
  \biggl[
    3 - \int_{z}^{\infty}\!
        \mathcal{J}_+(t)\,\mathrm{d}t
  \biggr]$
\\[1em]
$\omega_{\mathrm{eff}}\,\tau$ &
$\displaystyle
  -\frac{8\pi^2 i}{3}\,
  \frac{\varepsilon_-}{\varepsilon_+}\,
  \frac{m_e}{\varepsilon_+}\,
  \frac{(-\Delta)^{1/2}}{\omega}\;
  \mathcal{J}_+(z)$
\\[1em]
$\omega_{\mathrm{eff}}^2\,\tau^2$ &
$\displaystyle
  \frac{16\pi^2}{3}\,
  \frac{\varepsilon_-}{\varepsilon_+}\,
  \frac{(-\Delta)}{\omega}\,
  \frac{m_e^2}{\varepsilon_+^2}\;
  \mathcal{J}_-(z)$
\\[1em]
$\alpha^2$ &
$\displaystyle
  \frac{2\pi^2}{3}\,
  \frac{\varepsilon_-}{\varepsilon_+}\,
  \frac{(-\Delta)}{\omega}\,
  \frac{m_e^2}{\varepsilon_+^2}
  \biggl[
    3 - \int_{z}^{\infty}\!
        \mathcal{J}_+(t)\,\mathrm{d}t
    - \mathcal{J}_-(z)
  \biggr]$
\\[1em]
$\beta^2$ &
$\displaystyle
  \frac{2\pi^2}{3}\,
  \frac{\varepsilon_-}{\varepsilon_+}\,
  \frac{(-\Delta)}{\omega}\,
  \frac{m_e^2}{\varepsilon_+^2}
  \biggl[
    3 - \int_{z}^{\infty}\!
        \mathcal{J}_+(t)\,\mathrm{d}t
    - \mathcal{J}_-(z)
  \biggr]$
\\[1em]
\bottomrule
\end{tabular}
\end{table}

\bibliography{references}

\end{document}